\documentclass[prd,aps,preprint,superscriptaddress,onecolumn,tightenlines,nofootinbib,eqsecnum,preprintnumbers,longbibliography,12pt]{revtex4-1}
\pdfoutput=1
\usepackage{amsmath,latexsym,amssymb,graphicx,ifpdf,slashed,color,hyperref,url,cancel,lipsum}
\usepackage[T1]{fontenc}
\usepackage{relsize}
\hypersetup{colorlinks,citecolor= red ,linkcolor= niceblue, urlcolor=niceblue}
\definecolor{niceblue}{rgb}{0.1,0.2,0.6}

\begin{document}

\def\Carleton{Department of Physics, Carleton University, Ottawa, Ontario K1S 5B6, Canada}

\title{General Kinetic Mixing in Gauged $U(1)_{L_\mu-L_\tau}$ Model for\\ Muon $g-2$ and Dark Matter}
\author{Timothy Hapitas}
\email[Electronic address: ]{TimHapitas@cmail.carleton.ca}
\author{Douglas Tuckler}
\email[Electronic address: ]{dtuckler@physics.carleton.ca}
\author{Yue Zhang}
\email[Electronic address: ]{yzhang@physics.carleton.ca}

\affiliation{\Carleton}
\date{\today}

\begin{abstract}
The gauged $U(1)_{L_\mu-L_\tau}$ extension of the Standard Model is a very simple framework that can alleviate the tension in muon anomalous magnetic dipole moment, reinforced by the recent Fermilab measurement. We explore experimental probes of the $(g-2)_\mu$ target with a general treatment of kinetic mixing between the $Z'$ gauge boson and the photon. The physical value of the kinetic mixing depends on a free parameter of the model and energy scale of a process. We find neutrino constraints on the $(g-2)_\mu$ target including Borexino, CE$\nu$NS, and white dwarfs are sensitive to this freedom and can be lifted if the kinetic mixing lies in proximity of zero at low momentum transfer. As a further step, we explore $L_\mu-L_\tau$ charged dark matter with a thermal origin and show that the same scenario of kinetic mixing can relax existing direct detection constraints and predict novel recoil energy dependence in the upcoming searches. Future joint effort of neutrino and dark matter experiments and precision spectral measurement will be the key to test such a theory.
\end{abstract}

\maketitle
\tableofcontents

\section{Introduction}

The quest for physics beyond the Standard Model (SM) is the central focus of particle physics nowadays.
There exist a number of strong motivations for pursuing this endeavor, especially the phenomena established experimentally that the SM cannot adequately explain.
It is inspiring and important to explore potential theories and connections behind some of these puzzles. 

A long-standing puzzle at the precision frontier is the discrepancy between the experimental and theoretical values of the muon anomalous magnetic dipole moment $a_\mu \equiv (g-2)_\mu/2$. Earlier this year the Fermilab Muon $g-2$ Collaboration announced the latest measurement of the muon anomalous magnetic dipole moment~\cite{Abi:2021gix} and,
when combined with the previous result from Brookhaven~\cite{Bennett:2006fi}, exhibits a $4.2\sigma$ deviation from the SM prediction~\cite{Aoyama:2020ynm}
\begin{equation}\label{eq:amuTension}
\Delta a_\mu = a_\mu^\text{exp} - a_\mu^\text{SM} =  (251 \pm 59) \times 10^{-11} \ .
\end{equation}
Although this observation has not yet reached a $5\sigma$ significance, it could happen within the coming years with improved experimental statistics and refined theoretical calculations.
Given that gauge theories have played an instrumental role in the establishment of the SM, it is particularly attractive to address the $(g-2)_\mu$ tension by postulating new symmetries of nature.
A very simple consideration is to extend the SM with a gauged $U(1)_{L_\mu - L_\tau}$ symmetry, where exchange of the gauge boson $Z^\prime$ at one-loop level can shift the theoretical prediction of $(g-2)_\mu$ towards the measured value~\cite{Baek:2001kca}.
A number of existing and upcoming experiments can be used to constrain the relevant parameter space of the model~\cite{Alekhin:2015byh, Bauer:2018onh, Gninenko:2018tlp, Kahn:2018cqs, Altmannshofer:2019zhy}.

At the same time, a burning question at the cosmic frontier is the nature of dark matter (DM). 
If DM is a particle and interacts with SM particles through a new force beyond gravity, a large number of opportunities become available for probing it in laboratories.
With the introduction of the $U(1)_{L_\mu - L_\tau}$ gauge interaction, it is natural to further speculate on its role in the search for DM.
If nature does work this way, the $(g-2)_\mu$ results would be considered an indirect glimpse toward the dark side of the universe.
It also provides a strong motivation to explore other potential signatures of such a DM candidate.

In this work, we explore the phenomenology of the gauged $U(1)_{L_\mu - L_\tau}$ model.
We pay special attention to neutrino experiments that could probe the $(g-2)_\mu$ target parameter space, as well as the prospects of $L_\mu-L_\tau$ charged DM.
A common feature shared by neutrinos and DM in this model is that their low-energy scattering with human-made detectors must occur through 
a kinetic mixing between the $Z'$ gauge boson and the photon.
In general, the total kinetic mixing is given by the sum of a bare mixing parameter (not calculable) and radiative corrections, and is a function of the momentum transfer in a particular physical process.
As a result, the interpretation of experimental constraints and prospects strongly depends on the nature of the kinetic mixing. 
In the literature,
it was often assumed the total kinetic mixing is zero at energy scales much higher than the $\tau$ lepton mass~\cite{Kamada:2015era, Ibe:2016dir, Araki:2017wyg, Gninenko:2018tlp, Bauer:2018onh, Asai:2020qlp}
and a number of experimental constraints were derived based on this assumption.
A vanishing high-scale kinetic mixing, although may be realized in specific frameworks~\cite{Arguelles:2016ney, Gherghetta:2019coi}, is by no means general.

We revisit these constraints with a more general treatment of the kinetic mixing and find the asymptotic values of the kinetic mixing do matter a lot.
First, the constraint on the $(g-2)_\mu$ target obtained from solar neutrino measurements at the Borexino experiment can be lifted
if the kinetic mixing function is suppressed at low momentum transfer (well below the muon mass).
The same applies to the coherent elastic neutrino-nucleus scattering (CE$\nu$NS) experiments and white dwarf cooling.
Second, for $L_\mu-L_\tau$ charged DM with a thermal origin, 
 direct detection constraints on the DM mass scale are also sensitive to the nature of kinetic mixing.
We find that a new viable window of DM mass above GeV scale can be opened if the kinetic mixing is suppressed at low momentum transfer,
coinciding with the scenario where neutrino constraints are weakened. Notably, such a mass window has been considered excluded in the literature by assuming a vanishing kinetic mixing at high scales.
Moreover,  the DM candidate in such a mass window offers new predictions at future low-threshold direct detection experiments.
The corresponding DM-nucleus scattering features a novel shape of the recoil energy spectrum that will serve as a ``smoking-gun'' signature for testing the nature of DM in this model.

This article is organized as follows.
In Sec.~\ref{sec:model} we present the gauged $U(1)_{L_\mu - L_\tau}$ model and derive the momentum dependence of a general kinetic mixing.
It sets the stage for phenomenological discussions of experimental probes using neutrinos and DM in Sec.~\ref{sec:neutrino} and \ref{sec:DM}, respectively.
While Sec.~\ref{sec:neutrino} mainly argues that several neutrino constraints on the $(g-2)_\mu$ target can be lifted, in Sec.~\ref{sec:DM} we will show that relaxing the existing direct search constraints can lead to novel exciting prospects for future DM experiments.
The conclusions are drawn in Sec.~\ref{sec:conclusion}.

\section{Gauged $U(1)_{L_\mu - L_\tau}$ Model}\label{sec:model}
As the starting point, we present the minimal Lagrangian which extends the SM with a $U(1)_{L_\mu - L_\tau}$ gauge symmetry,
\begin{equation}\label{eq:Lag}
\begin{split}
\mathcal{L}_{L_\mu - L_\tau} = \ &\mathcal{L}_{\rm SM} - \frac{1}{4} Z^\prime_{\alpha\beta} Z'^{\alpha\beta}  + \frac{1}{2} m^2_{Z^\prime} Z^\prime_\alpha Z^{\prime \alpha} + \frac{\varepsilon_0}{2} Z^\prime_{\alpha\beta}  F^{\alpha\beta}  \\
 & + g_{\mu\tau} \big( \bar\mu \gamma^\alpha \mu - \bar\tau \gamma^\alpha \tau  
+  \bar\nu_\mu \gamma^\alpha P_L\nu_\mu -  \bar\nu_\tau \gamma^\alpha P_L \nu_\tau\big) Z'_\alpha  \ , \\
\end{split}
\end{equation}
where $Z^\prime$ is the gauge boson of the $U(1)_{L_\mu - L_\tau}$ symmetry under which the lepton doublet $L_\mu$ ($L_\tau$) carry a positive (negative) unit charge, and $g_{\mu\tau}$ is the corresponding gauge coupling.
We assume that the $U(1)_{L_\mu - L_\tau}$ symmetry is already Higgsed and the $Z'$ gauge boson is massive.
A renormalizable bare kinetic mixing parameter $\varepsilon_0$ has been introduced between the $Z'$ and photon field strengths.
It is a free parameter and not calculable within this model.
We do not address the origin of neutrino masses and mixing which requires further model building (see~\cite{Zhou:2021vnf} for recent studies).
The DM sector will be introduced later in Sec.~\ref{sec:DM}.


\subsection{The $(g-2)_\mu$ target}

\begin{figure}[t]
\centering
\includegraphics[width=0.618\textwidth]{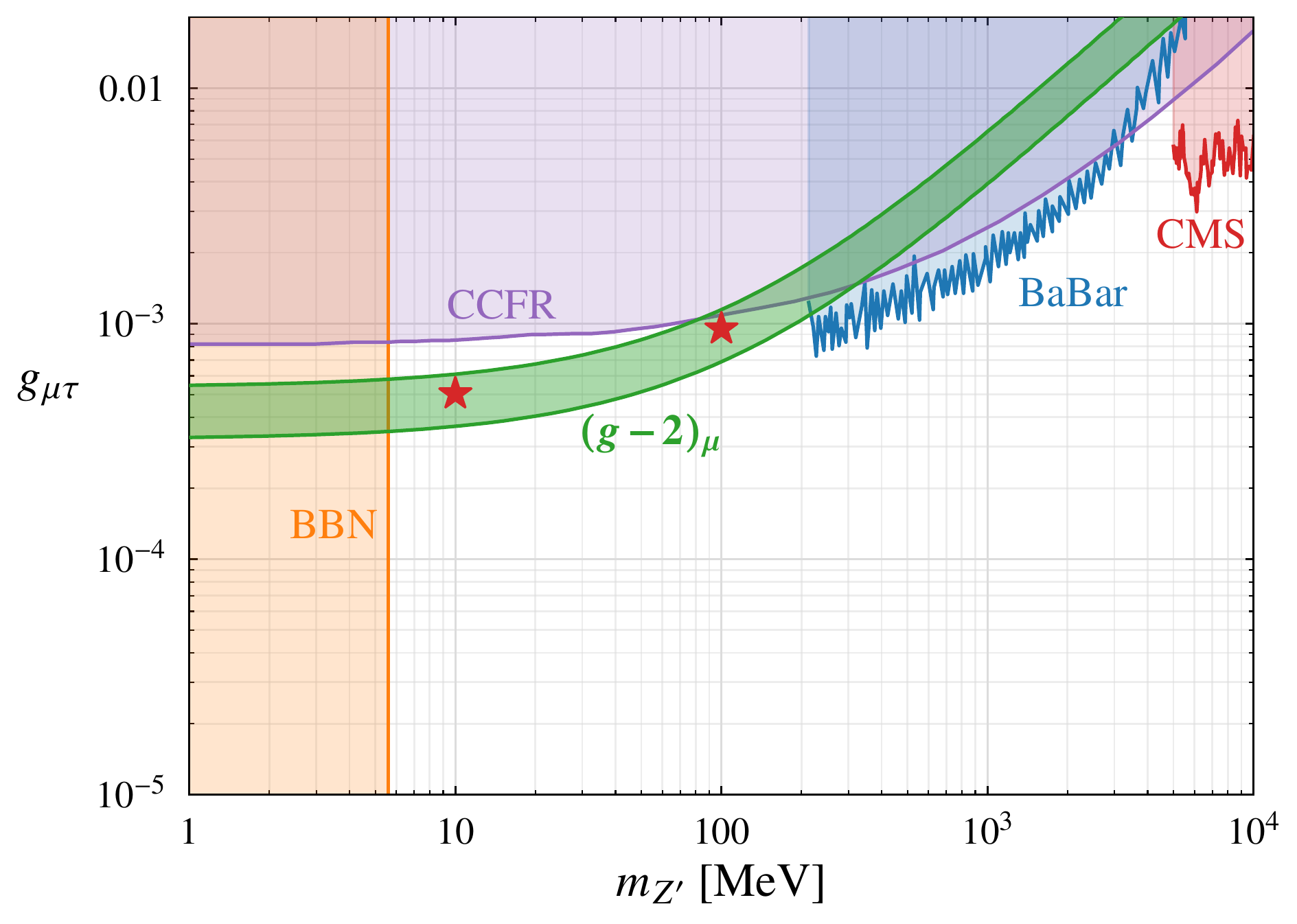}
\caption{Parameter space for addressing the $(g-2)_\mu$ tension in gauged $U(1)_{L_\mu - L_\tau}$ model (green band) shown along with experimental constraints that are independent of $Z'$-photon kinetic mixing. The exclusion regions include BBN (orange shaded region), neutrino trident search at CCFR (purple shaded region), BaBar search for $e^+e^-\to 4\mu$ (blue shaded region), and a CMS search for $p p \to \mu^+ \mu^- Z^\prime \to \mu^+ \mu^- \mu^+ \mu^- $ (red shaded region). The allowed mass range of the $Z'$ boson is $\sim 5-200$\,MeV. The two red five-stars represent the benchmark points to be used for the DM discussion in Sec.~\ref{sec:DM}.}\label{fig:aMu}
\end{figure}

It is well known that the virtual exchange of a $Z^\prime$ boson can induce a one-loop contribution to $a_\mu$~\cite{Baek:2001kca,Fayet:2007ua,Pospelov:2008zw,Davoudiasl:2014kua} given by
\begin{equation}\label{eq:aMu}
\Delta a_\mu = \frac{(g_{\mu\tau}+ e\varepsilon_0)^2}{4\pi^2} \int_0^1 dz \frac{m_\mu^2 z^2 (1-z)}{ m^2_{Z^\prime}(1-z)+m_\mu^2 z^2} \ ,
\end{equation}
With a vector current coupling to the muon, this contribution has the correct sign to shift the theoretical prediction of $a_\mu$ towards the experimentally measured value.
Throughout this work, we will restrict to the hierarchy where $\varepsilon_0$ is loop-factor suppressed compared to $g_{\mu\tau}/e$.
In this case, the $Z'$ couples just like a $U(1)_{L_\mu - L_\tau}$ gauge boson rather than a dark photon.\footnote{It is worth noting that the pure dark photon explanation of the $(g-2)_\mu$ has already been excluded~\cite{Alexander:2016aln, Fabbrichesi:2020wbt}, which suggests $\varepsilon_0$ cannot be much larger than $g_{\mu\tau}/e$.}
Neglecting the $\varepsilon_0$ piece of the contribution, Fig.~\ref{fig:aMu} depicts the parameter space in the $g_{\mu\tau}$ versus $m_{Z'}$ plane where the $(g-2)_\mu$ tension is alleviated, along with some existing constraints.
For $m_{Z^\prime} \gtrsim 5$ GeV the stongest constraints are from a CMS search for the $Z^\prime$ in the $p p \to \mu^+ \mu^- Z^\prime \to \mu^+ \mu^- \mu^+ \mu^-$ channel~\cite{CMS:2018yxg}.
The region where $Z'$ is heavier than twice the muon mass has been excluded by $e^+e^-\to 4\mu$ searches at BaBar~\cite{TheBABAR:2016rlg} and neutrino trident production searches at CCFR and CHARM~\cite{Altmannshofer:2014pba}, whereas the region with $Z'$ lighter than $\sim 6$\,MeV is excluded by the $\Delta N_{\rm eff}$ constraints from big-bang nucleosynthesis (BBN) and cosmic microwave background (CMB)~\cite{Ahlgren:2013wba,Kelly:2020pcy}.
The other constraints, such as that from the Borexino experiment, are not displayed here because of additional model dependence, as will be clarified below.~\footnote{The $\Delta N_{\rm eff}$ constrainted can be modified for a nonzero kinetic mixing but only for values of $g_{\mu\tau}$ much smaller than shown in Fig.~\ref{fig:aMu}~\cite{Escudero:2019gzq}.}
The viable mass window ($5\,{\rm MeV} \lesssim m_{Z'} \lesssim 200\,$MeV) for the $(g-2)_\mu$ explanation serves as a motivated target for many future experiments~\cite{Alekhin:2015byh, Bauer:2018onh, Gninenko:2018tlp, Kahn:2018cqs, Altmannshofer:2019zhy}.

\subsection{General kinetic mixing and momentum dependence}\label{subsec:running}

\begin{figure}[t]
\centering
\includegraphics[width=0.7\textwidth]{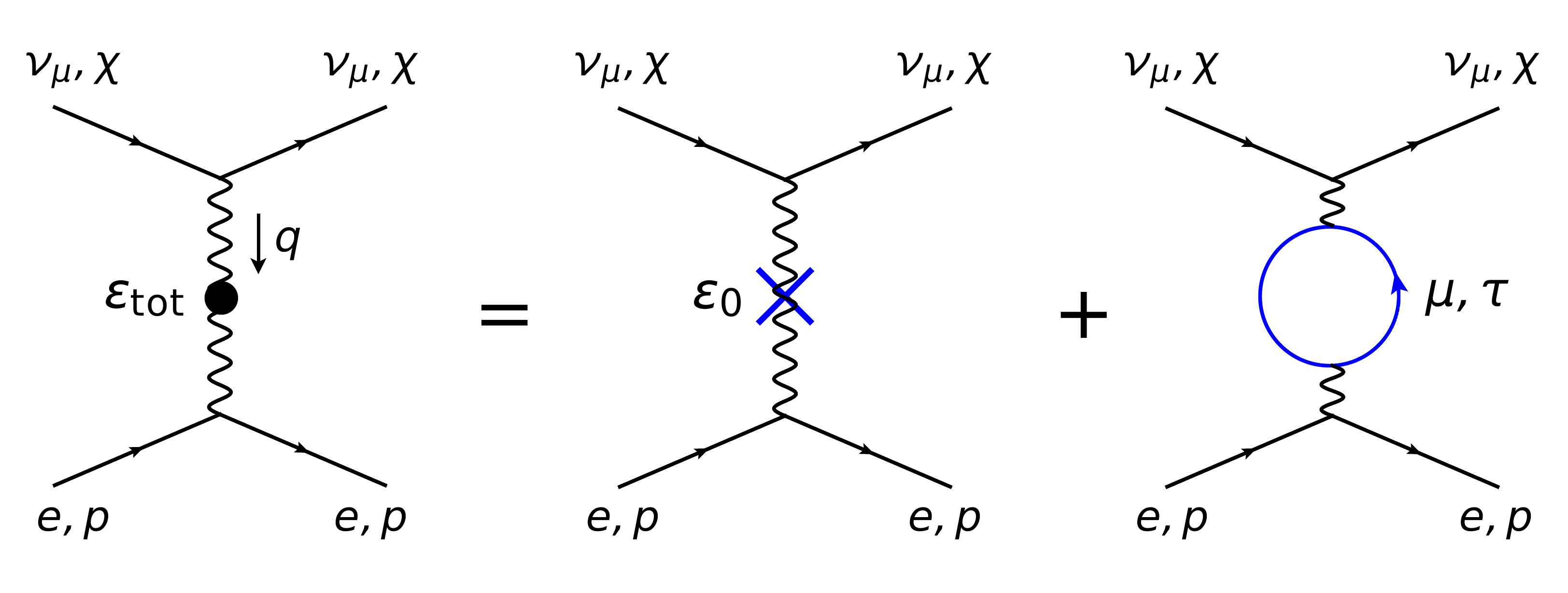}
\caption{Feynman diagrams for low energy neutrino or DM detection in gauged $U(1)_{L_\mu - L_\tau}$ model. The scattering with proton (or nucleus, electron) target happens through the kinetic mixing between the gauge boson $Z'$  and the photon. 
In general, the effective kinetic mixing depends on the momentum transfer $q^\mu$, or $Q = \sqrt{-q^2}$.}\label{fig:diagrams}
\end{figure}

The $Z'$ gauge boson of $U(1)_{L_\mu - L_\tau}$ does not directly couple to electrons or quarks. Therefore, the scattering of muon neutrinos or $L_\mu - L_\tau$ charged DM (to be introduced in Sec.~\ref{sec:DM}) in detectors must occur through the kinetic mixing between the $Z'$ and the photon (see Fig.~\ref{fig:diagrams}). The total kinetic mixing receives contributions from both the tree-level mixing term in Eq.~\eqref{eq:Lag} and radiative corrections with virtual $\mu$ and $\tau$ lepton exchange at loop level. It takes the following momentum transfer dependent form
\begin{equation}\label{eq:epsilontotal}
\varepsilon_{\rm tot} (Q) = \varepsilon_0  - \frac{e g_{\mu\tau}}{2\pi^2} \int_0^1 dx \, x(1-x) \log \left[ \frac{m_\tau^2 + x(1-x)Q^2}{m_\mu^2 + x(1-x)Q^2} \right] \ , 
\end{equation}
where $Q = \sqrt{-q^2}>0$, with $q^\mu$ being the four-momentum transfer in the scattering process. 
As mentioned earlier, the bare mixing parameter $\varepsilon_0$ is not calculable.

It is straightforward to verify that at zero or infinitely large momentum transfer
the total kinetic mixing approaches a constant. We denote these asymptotic values as $\varepsilon_{\rm IR} = \varepsilon_0 - \frac{e g_{\mu\tau}}{12\pi^2} \log \frac{m_\tau^2}{m_\mu^2}$ and $\varepsilon_{\rm UV} = \varepsilon_0$, respectively. Within the particle content of this model (introduced in Eq.~\eqref{eq:Lag}), they always satisfy, 
\begin{equation}
\varepsilon_{\rm UV} - \varepsilon_{\rm IR} = \frac{e g_{\mu\tau}}{12\pi^2} \log \frac{m_\tau^2}{m_\mu^2} \ .
\end{equation} 
Expanding to the next order, we find the following momentum transfer dependence
\begin{equation}\label{eq:twolimits}
\begin{split}
\varepsilon_{\rm tot} (Q) \simeq \varepsilon_{\rm IR} + \frac{e g_{\mu\tau}}{60\pi^2} \frac{Q^2}{m_\mu^2}\ , \quad \quad{\rm for}\ Q \ll m_\mu \ , \\
 \varepsilon_{\rm tot} (Q) \simeq \varepsilon_{\rm UV} - \frac{e g_{\mu\tau}}{2\pi^2} \frac{m_\tau^2}{Q^2}\ , \quad \quad{\rm for}\ Q \gg m_\tau \ .
\end{split}
\end{equation}
In contrast, for intermediate momentum transfer with $m_\mu < Q < m_\tau$, the $Q$ dependence is approximately logarithmic,
\begin{equation}
\frac{d\varepsilon_{\rm tot}}{d\log Q} \simeq \frac{e g_{\mu\tau}}{6\pi^2}  \ ,
\end{equation}
which corresponds to a running kinetic mixing.

\begin{figure}[t]
\centering
\includegraphics[width=0.618\textwidth]{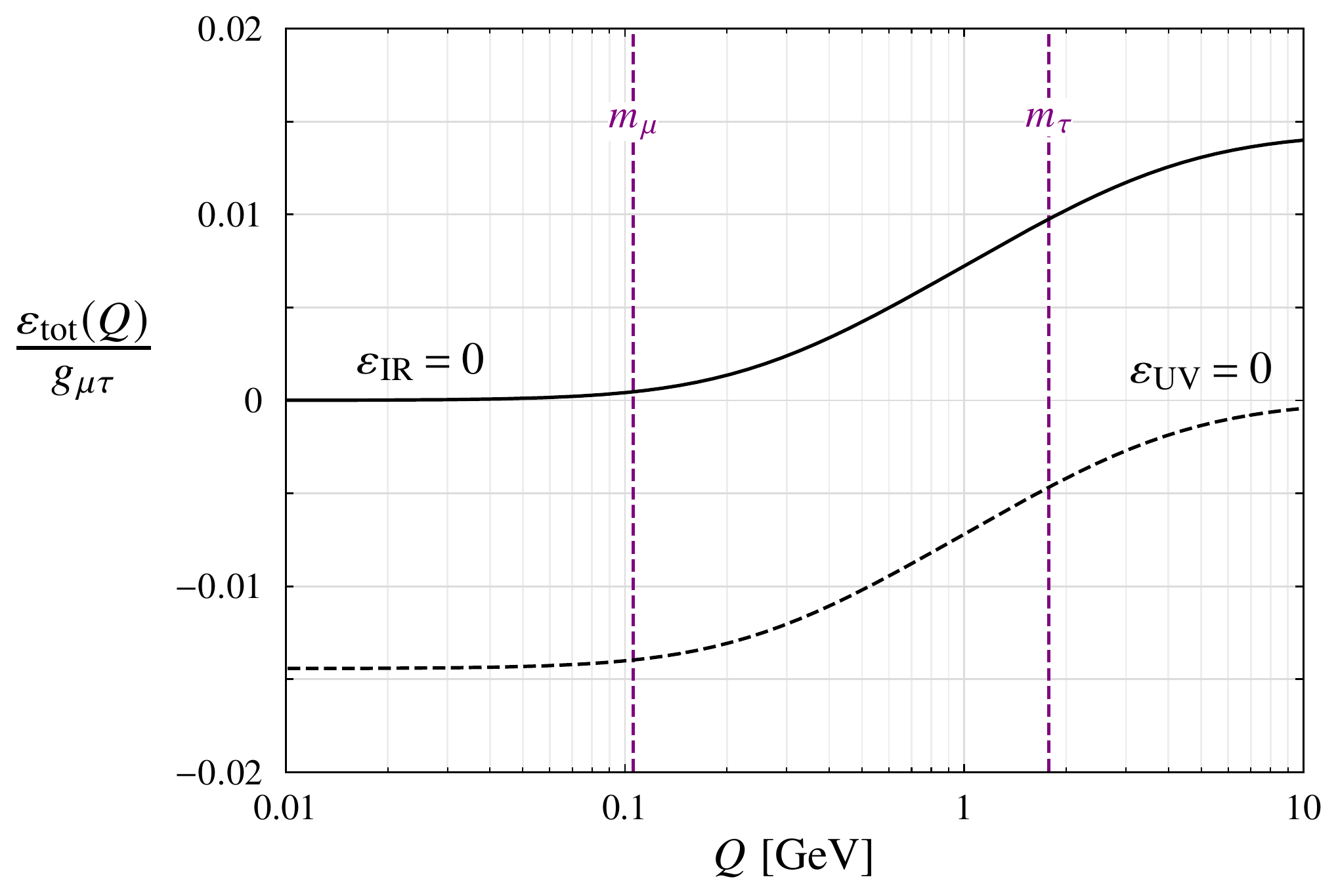}
\caption{Momentum dependent kinetic mixing $\varepsilon_\text{tot}(Q)$ as a function of momentum transfer $Q$. The solid (dashed) curve corresponds to the boundary condition $\varepsilon_\text{IR} = 0$ ($\varepsilon_\text{UV} = 0$). The vertical, dashed purple lines indicate where the momentum transfer is equal to $m_\mu$ and $m_\tau$. }\label{fig:running}
\end{figure}

This exercise shows that the exact form of $\varepsilon_{\rm tot}$ not only depends on the momentum transfer $Q$, but also its boundary conditions, $\varepsilon_{\rm IR}$ or $\varepsilon_{\rm UV}$.
This can be seen in Fig.~\ref{fig:running} which depicts $\varepsilon_{\rm tot}$ as a function of $Q$ for two distinct choices of boundary values.
In the case where $\varepsilon_{\rm IR} = 0$ (solid curve), $\varepsilon_{\rm tot}$ remains close to zero until $Q$ becomes larger than $m_\mu$. For $Q\gg m_\tau$,
$\varepsilon_{\rm tot}$ approaches $ \frac{e g_{\mu\tau}}{12\pi^2} \log ({m_\tau^2}/{m_\mu^2}) \simeq g_{\mu\tau}/70$.
In the second case where $\varepsilon_{\rm UV}=0$ (dashed curve), $\varepsilon_{\rm tot}$ remains close to zero for $Q\gg m_\tau$ and approaches $\varepsilon_{\rm IR}\simeq-g_{\mu\tau}/70$ for $Q\ll m_\mu$.
The latter corresponds to a rather common choice made in the literature~\cite{Kamada:2015era, Ibe:2016dir, Araki:2017wyg, Gninenko:2018tlp, Bauer:2018onh, Asai:2020qlp}.
More generally, the boundary conditions can be varied continuously such that they intersect the above two special scenarios.

Because our knowledge of higher scale physics is limited, in this work we keep an open mind to all possible boundary choices for the kinetic mixing.
Instead of judging which one is more appealing, we take a phenomenological approach and quantify their different implications in experiments where the gauged $U(1)_{L_\mu - L_\tau}$ model could be tested.

Here is an important observation that sets the stage for the remainder of this article.
In a number of neutrino and DM experiments, where interactions are mediated through the $Z'$-photon kinetic mixing,
the typical momentum transfer $Q$ lies well below the mass of the muon.
As a result, the two scenarios depicted Fig.~\ref{fig:running} can predict drastically different scattering rates and recoil energy spectra.
In the first case ($\varepsilon_{\rm IR} = 0$) the detection rates are often highly suppressed and the $Q$ dependence is important, whereas in the second case ($\varepsilon_{\rm UV}=0$) setting $\varepsilon_{\rm tot}\simeq-g_{\mu\tau}/70$ is a good approximation at low $Q < m_\mu$.
As we will see, these two choices lead to drastically different interpretations of neutrino and DM results, impacting the prospects of future experiments.

\section{Implication for Neutrino Probes of the $(g-2)_\mu$ Target}\label{sec:neutrino}

In this section, we revisit low-energy neutrino probes of the $(g-2)_\mu$ target in the context of the gauged $U(1)_{L_\mu - L_\tau}$ model. 
We derive constraints for the two kinetic mixing scenarios presented in Fig.~\ref{fig:running} of Sec.~\ref{subsec:running}.
The main point is that these experimental constraints are sensitive to the free parameter in the kinetic mixing and can be made substantially weaker than quoted in the literature.

\subsection{Borexino}\label{sec:borexino}

The Borexino experiment has obtained a measurement of solar neutrino-electron scattering. The result was originally shown to set a useful constraint on the gauged $U(1)_{B-L}$ model where the gauge boson directly couples to both electrons and neutrinos~\cite{Harnik:2012ni}. It was later reinterpreted in the context of the gauged $U(1)_{L_\mu - L_\tau}$ model~\cite{Kamada:2015era,Araki:2017wyg,Gninenko:2020xys}, where neutrino-electron scattering can occur through the $Z'$-photon kinetic mixing in addition to the weak interaction.
By assuming $\varepsilon_{\rm tot}\simeq-g_{\mu\tau}/70$, corresponding to the dashed curve in Fig.~\ref{fig:running}, an upper limit on the coupling $g_{\mu\tau}$ was derived as a function of the $Z'$ mass,
\begin{equation}\label{eq:Borexino}
|g_{\mu\tau}| < 6\times 10^{-4} \left( {m_{Z'}}/{10\,\rm MeV} \right) \ .
\end{equation}
Contrasting with the $(g-2)_\mu$ favored region, this sets the leading lower bound (around 10 MeV) on the $Z'$ mass. 

\begin{figure}[t]
\centering
\includegraphics[width=0.45\textwidth]{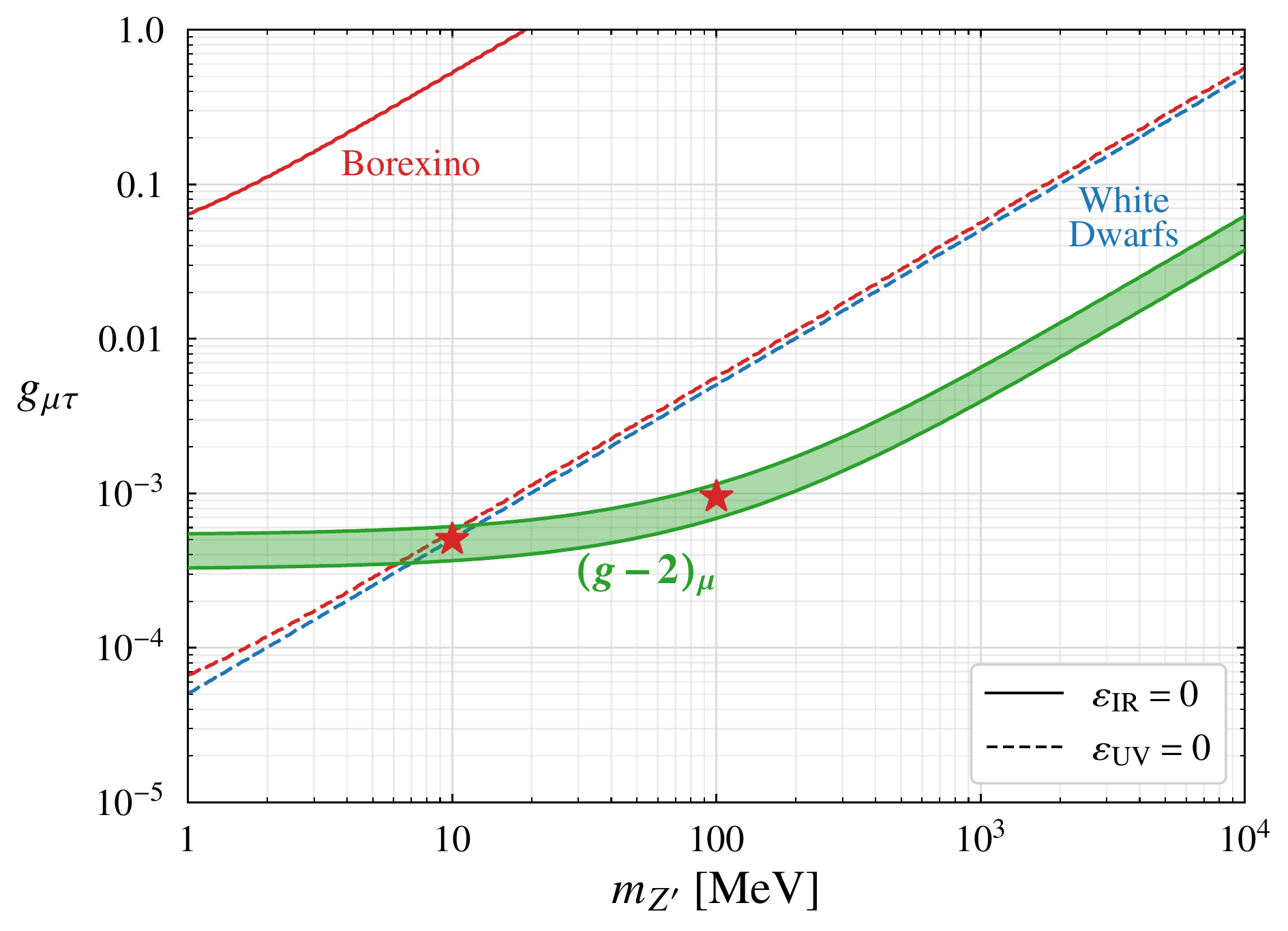} \\[4ex]
\includegraphics[width=0.45\textwidth]{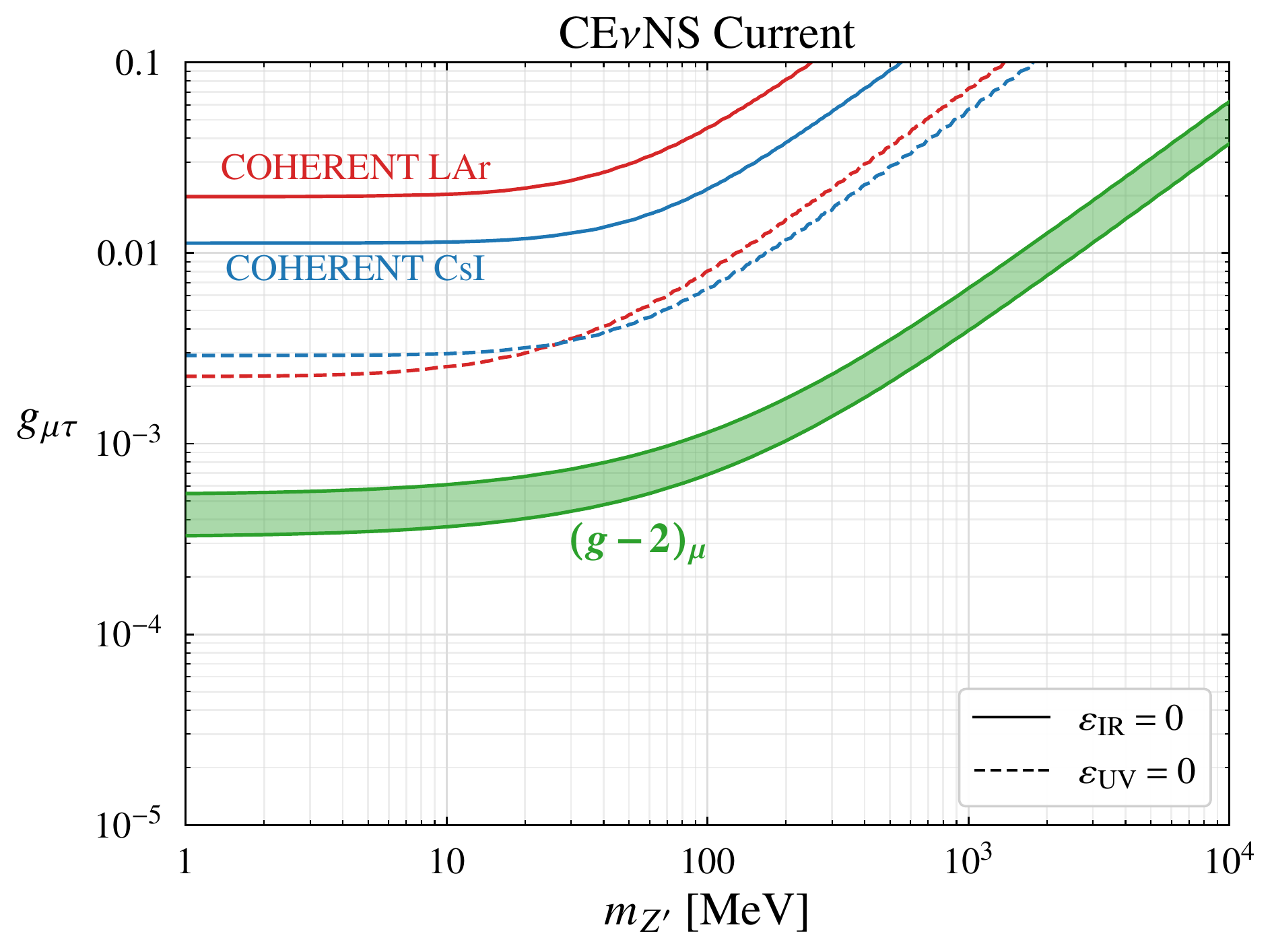}
\includegraphics[width=0.45\textwidth]{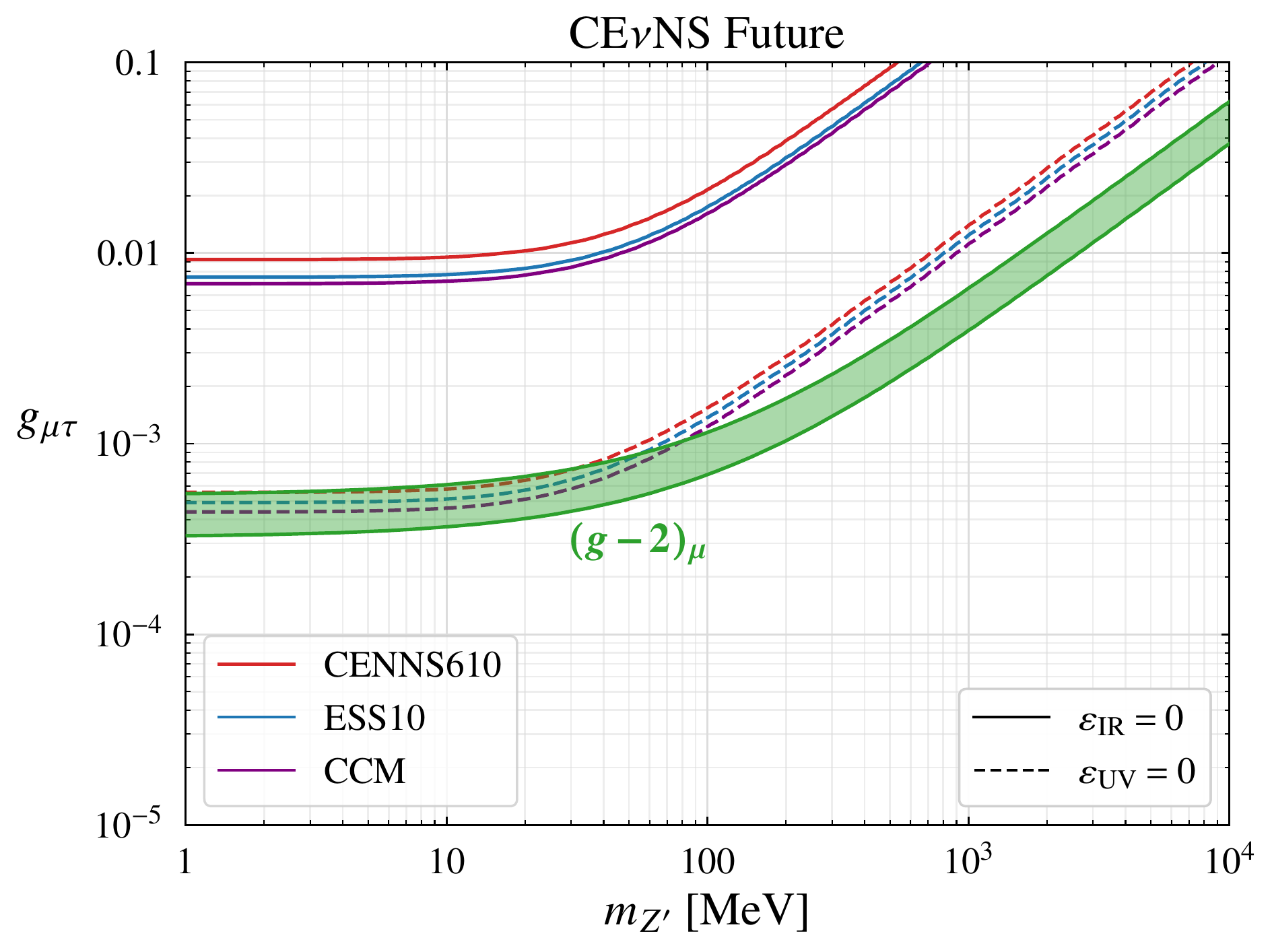}
\caption{Current and future reach (upper bound) on $g_{\mu\tau}$ from neutrino experiments, including Borexino and white dwarfs (first row) and the CE$\nu$NS experiments (second row).
The dashed curves correspond to the kinetic mixing scenario with $\varepsilon_{\rm UV}=0$, and the solid curves correspond to the kinetic mixing scenario with $\varepsilon_{\rm IR}=0$ which are much weaker.}\label{fig:neutrino}
\end{figure}

However, this constraint is model dependent due to the ambiguity of the kinetic mixing. As discussed in Sec.~\ref{subsec:running},
the general kinetic mixing at Borexino takes the asymptotic form
\begin{equation}
\varepsilon_{\rm tot}  \simeq \varepsilon_{\rm IR} + \frac{e g_{\mu\tau}}{60\pi^2} \left(\frac{{\rm MeV}}{m_\mu}\right)^2 \ ,
\end{equation}
where we have used the fact that the typical momentum transfer in solar neutrino-electron scattering process is around the MeV scale, much lower than the muon mass.
Thus the first line of Eq.~\eqref{eq:twolimits} applies.
Setting the boundary value $\varepsilon_{\rm IR}=0$, which corresponds to the solid curve in Fig.~\ref{fig:running}, we obtain
\begin{equation}
\varepsilon_{\rm tot}  \simeq 5\times10^{-8} g_{\mu\tau} \ .
\end{equation}
Compared to the other choice $\varepsilon_{\rm tot}\simeq-g_{\mu\tau}/70$, the upper bound Eq.~\eqref{eq:Borexino} can be relaxed by a factor of $\sim 500$,
making it irrelevant for probing the $(g-2)_\mu$ target. This can be seen in the top plot of Fig.~\ref{fig:neutrino} where the constraints from Borexino are given by the solid and dashed red curves for $\varepsilon_{\rm IR}=0$ and $\varepsilon_\text{UV}=0$, respectively. The bounds on $g_{\mu\tau}$ are placed using measurements of the $^7$Be neutrino line, which has the most precisely measured flux~\cite{Borexino:2017rsf}. The bound is significantly weaker for the boundary condition $\varepsilon_{\rm IR}=0$.
This simple exercise demonstrates that caution should be taken on constraints from low energy neutrino scattering experiments in the gauged $U(1)_{L_\mu - L_\tau}$ model. They are not as robust as those that are independent of the kinetic mixing (as shown in Fig.~\ref{fig:aMu}).

\subsection{CE$\nu$NS}

Similar to neutrino-electron scattering, the $Z^\prime$ of $U(1)_{L_\mu - L_\tau}$ can contribute to the CE$\nu$NS processes. A first measurement of this rare process was made by the COHERENT experiment using a CsI target~\cite{Akimov:2017ade, Akimov:2018vzs} and later using a liquid argon (LAr) target~\cite{COHERENT:2020iec,COHERENT:2020ybo}. These results are compatible with SM predictions and can be used to constrain the parameter space of the $U(1)_{L_\mu - L_\tau}$ model.

In CE$\nu$NS experiments, neutrinos are produced from pions that decay at rest, with energies below the muon mass. The neutrino flux consists of a prompt component of muon-neutrinos from $\pi^+ \to \mu^+ \nu_\mu$, and a delayed component of muon anti-neutrinos and electron neutrinos produced from the subsequent decay of the muon. The neutrino spectra seen in the detector are~\cite{Akimov:2017ade, Akimov:2018vzs}
\begin{equation}
\begin{split}
\frac{dN_{\bar{\nu}_\mu}}{dE_\nu} &= \eta \frac{64 E^2_\nu}{m^3_\mu} \bigg( \frac{3}{4} - \frac{E_\nu}{m_\mu}\bigg) \ ,\\
\frac{dN_{\nu_e}}{dE_\nu} &= \eta  \frac{192 E^2_\nu}{m^3_\mu} \bigg( \frac{1}{2} - \frac{E_\nu}{m_\mu}\bigg) \ , \\
\frac{dN_{\nu_\mu}}{dE_\nu} &= \eta \delta \bigg( E_\nu - \frac{m_\pi^2 - m_\mu^2} {2m_\pi}\bigg) \ ,
\end{split}
\end{equation}
where $\eta = r N_\text{POT}/4\pi L^2$ is a normalization factor that depends on the experimental design, with $r$ being the number of neutrinos of a particular flavor produced per proton-on-target (POT) and $L$ the distance between the neutrino source and the detector. Here we will consider constraints from existing COHERENT CsI and LAr experiments, as well as projections for future LAr experiments. In particular, we will consider a future 610 kg experiment CENNS-610 \cite{COHERENT:2019kwz}, a proposed 10-kg experiment at the European Spallation Source (ESS) \cite{Baxter:2019mcx}, and the 7-ton fiducial mass Coherent CAPTAIN-Mills (CCM) experiment \cite{CCM,CCM:2021leg}. Details of the experimental parameters can be found in Table~\ref{tab:coherent} in the appendix.

The number of expected CE$\nu$NS events in a given detector can be calculated as
\begin{equation}\label{eq:Ncevns}
N_\text{CE$\nu$NS} = \sum_{\nu_\alpha} N_\text{targ} \int_{E_{\rm th}}^{E_R^{\rm max}} dE_R \int_{E_\nu^{\rm min}}^{E_\nu^{\rm max}} dE_\nu \frac{dN_{\nu_\alpha}}{dE_\nu} \epsilon(E_R) \frac{d\sigma_{\nu_\alpha A}}{dE_R}  \ ,
\end{equation}
where $E_\nu^{\rm min}=\sqrt{M_A E_R/2}$, $E_\nu^{\rm max}=m_\mu/2$, $E_R^{\rm max} = m_\mu^2/(2M_A)$, and $M_A$ is the target nucleus mass.
The function $\epsilon(E_R)$ is the efficiency factor, and the differential cross section takes the form
\begin{equation}
\frac{d\sigma_{\nu_\mu N}}{dE_R} = \frac{d\sigma_{\bar\nu_\mu N}}{dE_R} = \frac{m_A}{2\pi} \left( 1 - \frac{M_A E_R}{2 E_\nu^2} \right) 
\left| \frac{G_F Q_{\nu A}}{2} + \frac{g_{\mu\tau} \varepsilon_{\rm tot} Z e}{2 m_A E_R + m_{Z'}^2} \right|^2 F_{\rm Helm}^2(E_R) \ ,
\end{equation}
where $Q_{\nu A} = A - (1-4\sin^2\theta_W) Z$.
Here, the sign of $g_{\mu\tau}$ matters and for illustration we consider $g_{\mu\tau}>0$.
In Eq.~\eqref{eq:Ncevns}, for $\alpha =e$, the leading cross section is SM-like. The efficiency factor $\epsilon (E_R)$ for the CsI is taken from Ref.~\cite{Akimov:2018vzs}, while for all LAr experiments we adopt the efficiency factor from Ref.~\cite{COHERENT:2020ybo}.
The Helm form factor takes the form~\cite{Helm:1956zz,Duda:2006uk,Jungman:1995df}
\begin{equation}\label{eq:helm}
F(E_R) = \frac{3j_1( Q r_n)}{Q r_n} \exp(-Q^2 s^2/2) \ ,
\end{equation}
where $Q=\sqrt{2M_A E_R}$, $r_n=1.14 A^{1/3}\,{\rm fm}$, and $s \simeq 0.9\,$fm.

To determine the upper bounds on $g_{\mu\tau}$ from current and future experiments we use a chi-square test, which is detailed in Appendix~\ref{sec:stats}. In the bottom row of Fig.~\ref{fig:neutrino}, we show the current (left plot) and future (right plot) constraints on the parameter space of the $U(1)_{L_\mu - L_\tau}$ gauge boson. In both plots, the dashed curves correspond to the boundary condition $\varepsilon_\text{UV} = 0$, while the solid curves correspond to the boundary condition $\varepsilon_\text{IR} = 0$. Clearly, the choice of boundary condition impacts the constraints strongly. 

For $\varepsilon_\text{UV} = 0$, the current experiments do not yet constrain the favoured $(g-2)_\mu$ parameter space (green shaded band), but the future experiments will begin probing this target for $m_{Z^\prime} \lesssim 100$ MeV. On the other hand, for $\varepsilon_\text{IR} = 0$, all the constraints weaken substantially. In the bottom-right plot of Fig.~\ref{fig:neutrino}, even the sensitivity of future experiments considered above are orders of magnitude too weak. In this case, the parameter space that explains the $(g-2)_\mu$ anomaly will remain in tact after those experiments. 

\subsection{White dwarf}

Here we briefly comment on the constraint from white dwarf cooling \cite{Bauer:2018onh, Dreiner:2013tja}. The gauged $U(1)_{L_\mu - L_\tau}$ model  provides a novel cooling process that could cause excessive energy loss of the star: plasmon decaying into $\nu_\ell\bar\nu_\ell$ ($\ell=\mu, \tau$) through the exchange of the $Z'$ boson. Because the temperature of white dwarfs ($\sim 10^7\,$K $\sim $ keV) is much lower than the $Z'$ mass scale of interest to $(g-2)_\mu$, it is sufficient to consider the effective Lagrangian, 
\begin{equation}
\mathcal{L}_{\rm eff} =({g_{\mu\tau} \varepsilon_{\rm tot} e}/{m_{Z'}^2})(\bar e \gamma^\alpha e) (\bar \nu_\ell \gamma_\alpha P_L \nu_\ell) \ .
\end{equation}
With the rather low momentum transfer, it is clear that choosing a kinetic mixing with $\varepsilon_{\rm IR}=0$ leads to a much weaker constraint than the $\varepsilon_{\rm UV}=0$ case, by a factor of $\sim 6\times 10^5$.
This is derived using a similar approach as in Sec.~\ref{sec:borexino}.~\footnote{For white dwarf cooling, we have $q^2>0$ where the $q$ is the four momentum transfer through the virtual $Z'$ to final state $\nu\bar\nu$. However, the momentum dependence derived in Eq.~\eqref{eq:twolimits} still holds after an analytic continuation to the $Q^2=-q^2<0$ regime.}
The corresponding constraint curve for the $\varepsilon_{\rm IR}=0$ case lies outside (above) the range of $g_{\mu\tau}$ plotted in Fig.~\ref{fig:neutrino}.

We close this section by stressing once again that the momentum dependence in the $Z'$-photon kinetic mixing can strongly impact experimental constraints 
in the gauged $U(1)_{L_\mu - L_\tau}$ model. The model features an extra parameter --- the boundary condition of the kinetic mixing.
We have shown that low energy neutrino processes can be suppressed if the kinetic mixing is in the proximity of zero at low momentum transfer below the muon mass.
In contrast, constraints that do not involve the kinetic mixing still apply, including those from CMS, BaBar, CCFR, and BBN shown in Fig.~\ref{fig:aMu}.
The freedom of choosing a kinetic mixing value other than $\varepsilon_{\rm tot}=-g_{\mu\tau}/70$ has been considered in some previous studies~\cite{Banerjee:2018mnw, Escudero:2019gzq, Amaral:2021rzw},
but we think our work is the first to take into account the momentum transfer dependence when deriving the above experimental constraints.

\section{Implication for $L_\mu - L_\tau$ Charged Dark Matter}\label{sec:DM}

In this section, we go beyond the minimal gauged $U(1)_{L_\mu - L_\tau}$ model of Eq.~(\ref{eq:Lag}) by further considering that the $Z'$ gauge boson serves as a mediator between the visible sector and DM. This is a natural and well-motivated idea to pursue given the significance of existing $(g-2)_\mu$ tension. While there could be various incarnations for $L_\mu - L_\tau$ charged DM~\cite{Altmannshofer:2016jzy, Kamada:2018zxi, Foldenauer:2018zrz, Asai:2020qlp,Borah:2021jzu,Holst:2021lzm,Drees:2021rsg}, here we consider a rather minimal setup where 
the DM candidate $\chi$ is a vector-like (Dirac) fermion. The Lagrangian is
\begin{equation}\label{eq:LagDark}
\begin{split}
\mathcal{L} = \mathcal{L}_{L_\mu - L_\tau}  + i\bar \chi  \gamma^\alpha \partial_\alpha \chi + g_\chi \bar\chi \gamma^\alpha \chi \hat{Z}^\prime_\alpha - m_\chi \bar \chi \chi \ , \\
\end{split}
\end{equation}
where $g_\chi$ is the $U(1)_{L_\mu-L_\tau}$ gauge coupling for DM. In general, it is different from $g_{\mu\tau}$ introduced earlier, which acknowledges that the $L_\mu - L_\tau$ charge of $\chi$ can be different from those of $\mu$ and $\tau$ leptons.

We will explore the DM phenomenology in this model by always making the following assumptions:
\begin{itemize}
\item Throughout the analysis, we always assume that the $(g-2)_\mu$ anomaly is addressed by the $Z'$ boson, which fixes the coupling $g_{\mu\tau}$ as a function of $m_{Z'}$. 
\item We will assume that DM is a thermal relic of the early universe, which fixes the DM-$Z'$ coupling $g_\chi$ as a function of DM mass $m_\chi$.
\end{itemize}


\subsection{The thermal dark matter target}

We consider the DM candidate $\chi$ with a thermal origin, obtaining the observed relic abundance via the freeze out mechanism.
There are two classes of channels that can fix the relic abundance of DM today: annihilation into $Z'$ bosons, or annihilation into SM leptons via off-shell $Z'$ exchange.
We continue to assume the hierarchy, $g_{\mu\tau} \gg e\varepsilon_0$, so that the gauge coupling $g_{\mu\tau}$ plays the dominant role over the kinetic mixing.
The corresponding annihilation cross sections are
\begin{equation}\label{annihilationchannels}
\begin{split}
(\sigma v)_{\chi\bar\chi\to Z'Z'} &= \frac{g_\chi^4}{16\pi m_\chi^2} \left( 1 - \frac{m_{Z'}^2}{m_\chi^2} \right)^{3/2} \left( 1 - \frac{m_{Z'}^2}{2m_\chi^2} \right)^{-2} \ , \\
(\sigma v)_{\chi\bar\chi\to \nu_\ell \bar\nu_\ell} &= \frac{g_\chi^2 g_{\mu\tau}^2 m_\chi^2}{2\pi [(4m_\chi^2 - m_{Z'}^2)^2 + m_{Z'}^2 \Gamma_{Z'}^2]} \ , \\
(\sigma v)_{\chi\bar\chi\to \ell^+\ell^-} & = \frac{g_\chi^2 g_{\mu\tau}^2 m_\chi^2}{\pi [(4m_\chi^2 - m_{Z'}^2)^2 + m_{Z'}^2 \Gamma_{Z'}^2]} \left( 1 + \frac{m_{\ell_a}^2}{2m_\chi^2} \right) \sqrt{1 - \frac{m_{\ell_a}^2}{m_\chi^2}} \ ,
\end{split}
\end{equation}
where the flavor index $\ell=\mu$ or $\tau$. The $Z'$ decay width is~\cite{Kelly:2020pcy}
\begin{equation}
\Gamma_{Z'} = \frac{g_{\mu\tau}^2 m_{Z'}}{12\pi} \left[ 1 + \sum_{\alpha=\mu}^\tau (1+ 2 r_\alpha) (1-4r_\alpha)^{1/2} \Theta(1-4r_\alpha) \right] \ ,
\end{equation}
where $r_\alpha = m_\alpha^2/m_{Z'}^2$ and $\Theta$ is the Heaviside theta function.
 
\begin{figure*}[t]
\centering
\includegraphics[width=0.618\textwidth]{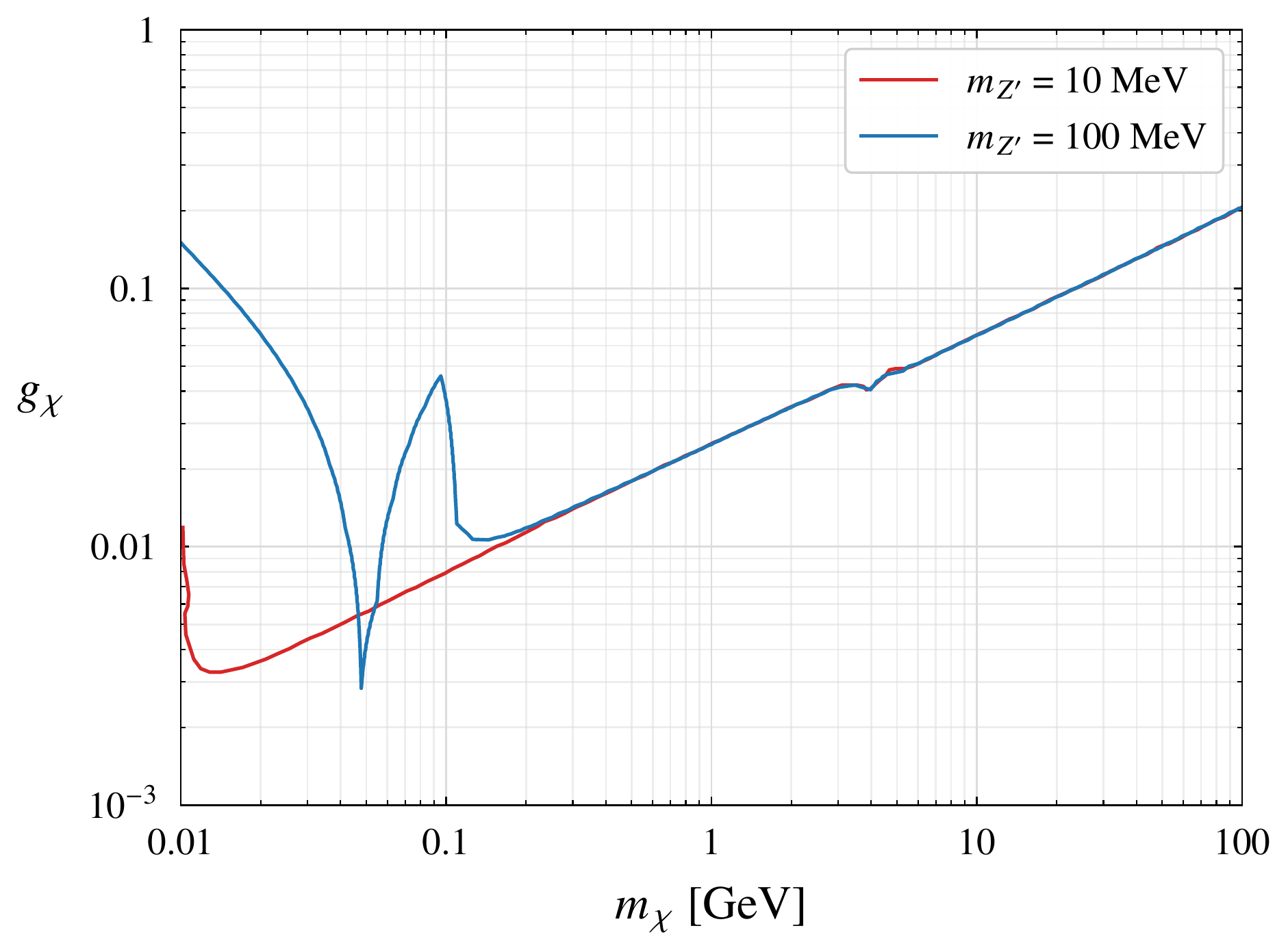}
\caption{Parameter space of DM in the $g_\chi$ versus $m_\chi$ plane that depicts where the observed relic density is successfully explained, for $m_{Z'} = $ 10 MeV (red curve) and $m_{Z'} = $ 100 MeV (blue curve). For $m_\chi \gg 100\,$MeV, the annihilation channel $\chi\bar\chi\to Z'Z'$ plays the dominant role. 
The kink on the curves around $m_\chi\simeq 4$\,GeV is mainly due to the large number of degree of freedom change when the universe was at the QCD phase transition temperature.}\label{fig:relic}
\end{figure*}

In Fig.~\ref{fig:relic}, the solid red and blue curves depict the parameter space where the DM relic density is successfully produced.
In this work, we are primarily interested in the direct detection prospects of DM in traditional (large) detectors, thus
the focus is on DM heavier than several hundred MeV.
Numerically, we find that the correct relic density favors $g_\chi \gg g_{\mu\tau}$ where the latter is used to fix the $(g-2)_\mu$ anomaly.
Therefore, the first annihilation channel in Eq.~\eqref{annihilationchannels} dominates the thermal freeze out.
Roughly, for $m_\chi\gg m_{Z'}$: 
\begin{equation}
g_\chi \simeq 0.025 \sqrt{\frac{m_\chi}{\rm GeV}} \ .
\end{equation}
%

\subsection{Upper limit on dark matter mass from direct detection}

There is a strong interplay between thermal relic density and DM direct detection in this model. Since the DM does not couple directly to quarks, its scattering with nuclei in a detector must occur through the $Z'$-photon kinetic mixing and is thus subject to similar model dependence as the neutrino scattering experiments discussed in Sec.~\ref{sec:model} and \ref{sec:neutrino}. With $g_{\mu\tau}$ fixed such that the $(g-2)_\mu$ anomaly is explained, the DM scattering rate is controlled by the coupling $g_\chi$ and the choice of kinetic mixing.

At the nucleon level, the low-energy DM scattering cross section takes the form
\begin{equation}
\label{eq:directdetectioncrosssection}
\sigma_{\chi p} \approx \frac{ g_\chi^2\varepsilon_{\rm tot}^2 e^2 \mu_{\chi p}^2}{\pi m_{Z'}^4} \ ,
\end{equation}
where $\mu_{\chi p} = m_\chi m_p/(m_\chi + m_p)$ is the reduced mass of DM-proton system.
An approximation has been made by neglecting the momentum transfer in the $Z'$ propagator, which is potentially important for light $Z'$.

To account for the momentum transfer dependence, we consider the nucleus level DM scattering rate derived using the standard halo model
\begin{equation}\label{eq:rate}
R = \frac{N_A M_A \rho_\odot g_\chi^2 Z^2 e^2}{2\pi m_\chi} \int_{E_{\rm th}}^{E_R^{\rm max}} d E_R \frac{\varepsilon^2_{\rm tot}(Q) F_{\rm Helm}^2(Q)}{(Q^2 + m_{Z'}^2)^2} \int_{v_{\rm min}(E_R)}^{v_{\rm esc}} d^3 v \frac{f(\vec{v})}{v} \ ,
\end{equation}
where $N_A$ is the number of target nucleus, $M_A$ and $Ze$ are the mass and electric charge of the nucleus,
$F_{\rm Helm}$ is the Helm form factor, and $Q \simeq \sqrt{2M_AE_R}$ for non-relativistic scattering.
The local DM mass density is $\rho_\odot=0.3\,{\rm GeV/cm^3}$ and the velocity distribution is Maxwellian
$f(\vec{v}) = C \exp(- |\vec{v} + \vec{v}_\odot|^2/v_0^2)\, \Theta(v_{\rm esc} - |\vec{v} + \vec{v}_\odot|)$,
where $\vec{v}$ is the DM velocity in the rest frame of the solar system, $v=|\vec{v}|$, $v_\odot=220\,{\rm km/s}$, $v_0=235\,{\rm km/s}$, $v_{\rm esc}=550\,{\rm km/s}$~\cite{Lin:2019uvt}, and $C$ is a normalization factor such that $\int d^3 v f(\vec{v})=1$. The Helm form factor is given by Eq.~\eqref{eq:helm}.
The limits of integration over velocity and recoil energy in Eq.~(\ref{eq:rate}) are
\begin{equation}
v_{\rm min}(E_R)= \sqrt{\frac{M_A E_R}{2 \mu_{\chi A}^2}}  \ , \quad\quad E_R^{\rm max} = \frac{2 \mu_{\chi A}^2 (v_{\rm esc} + v_\odot)^2}{M_A} \ ,
\end{equation}
where $\mu_{\chi A} = m_\chi M_A/(m_\chi + M_A)$ is the reduced mass of DM-nucleus system, and $E_{\rm th}$ is the energy threshold of the detector under consideration.

\begin{figure}[t]  
\includegraphics[width=0.45\textwidth]{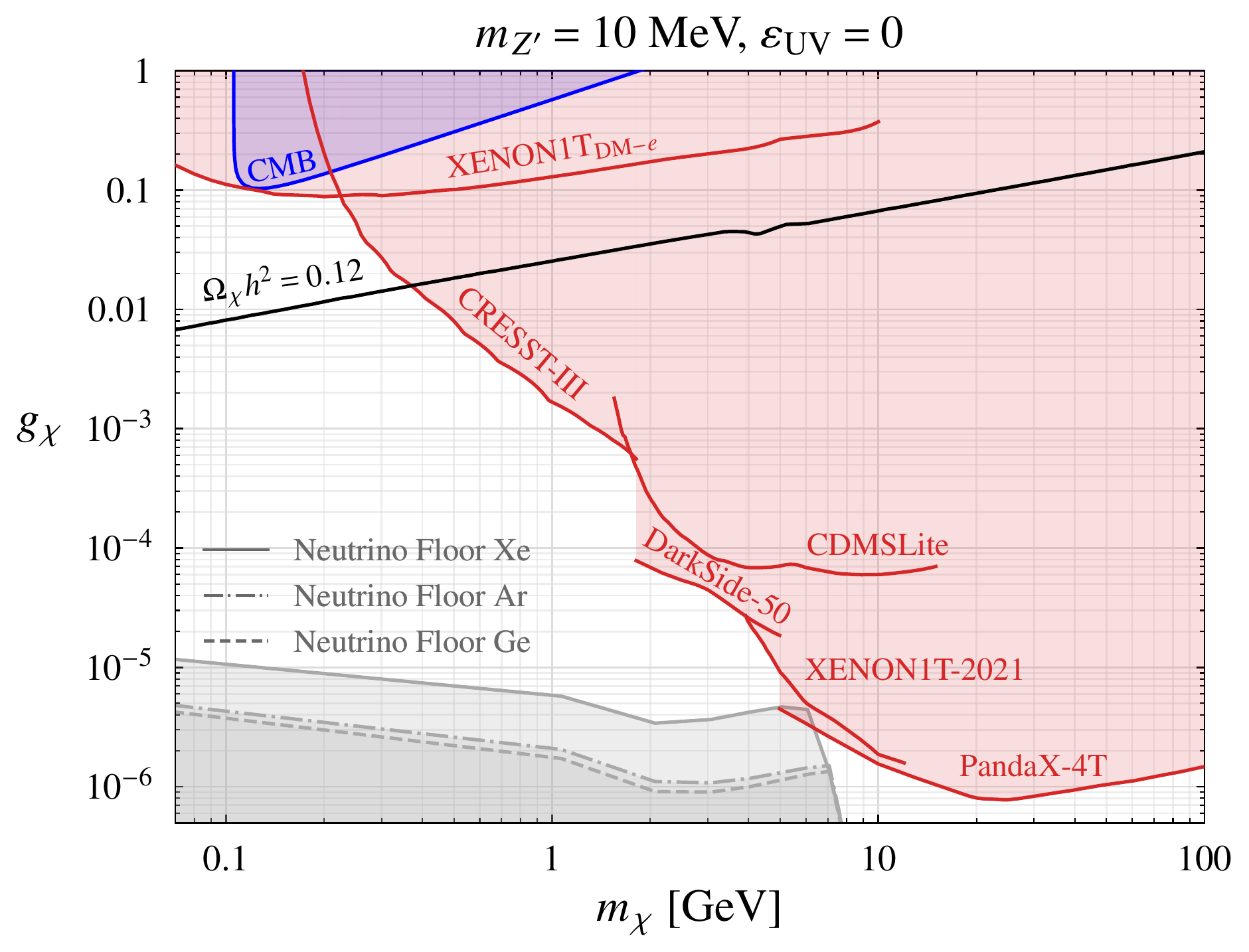}
\includegraphics[width=0.45\textwidth]{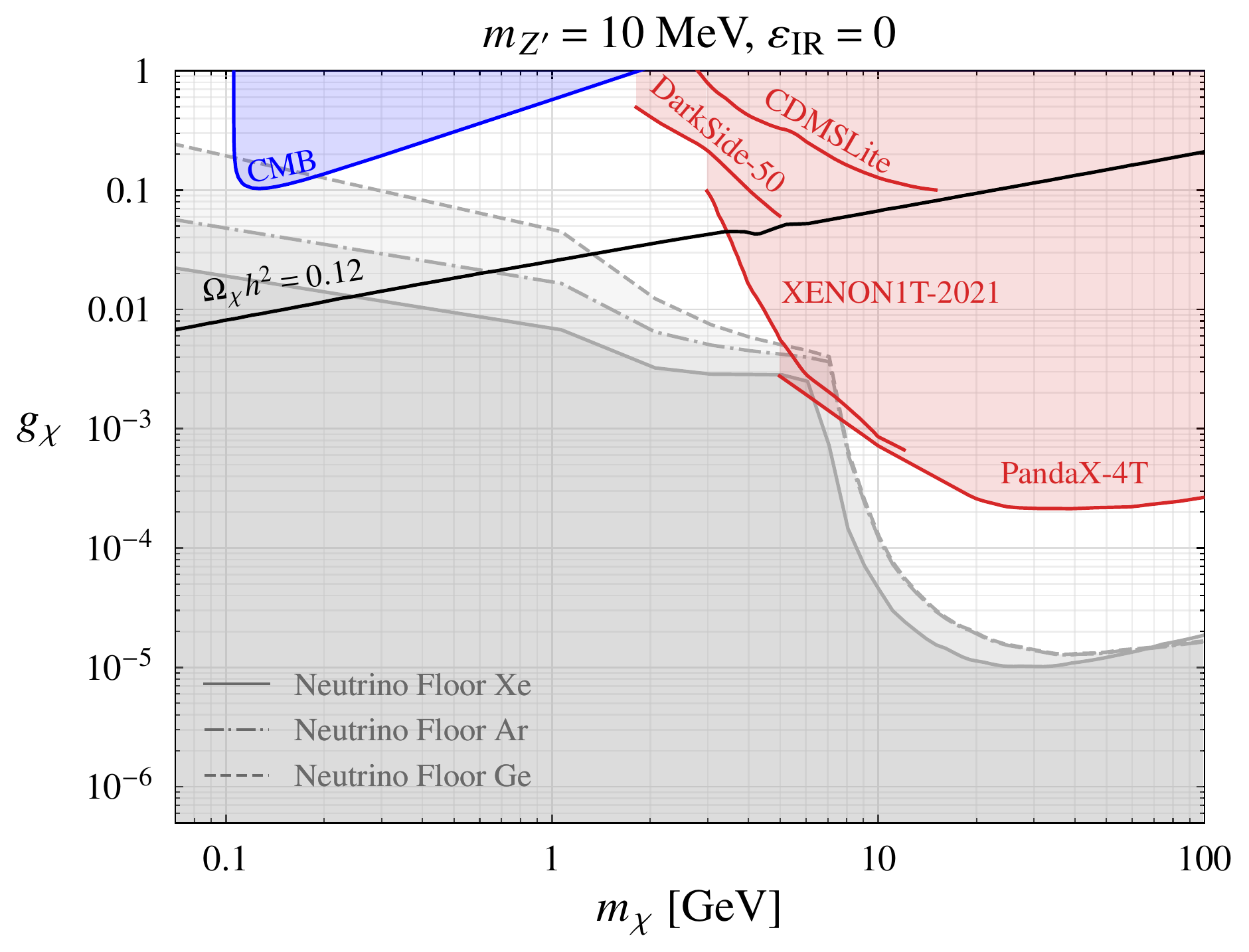}\\[2ex]
\includegraphics[width=0.45\textwidth]{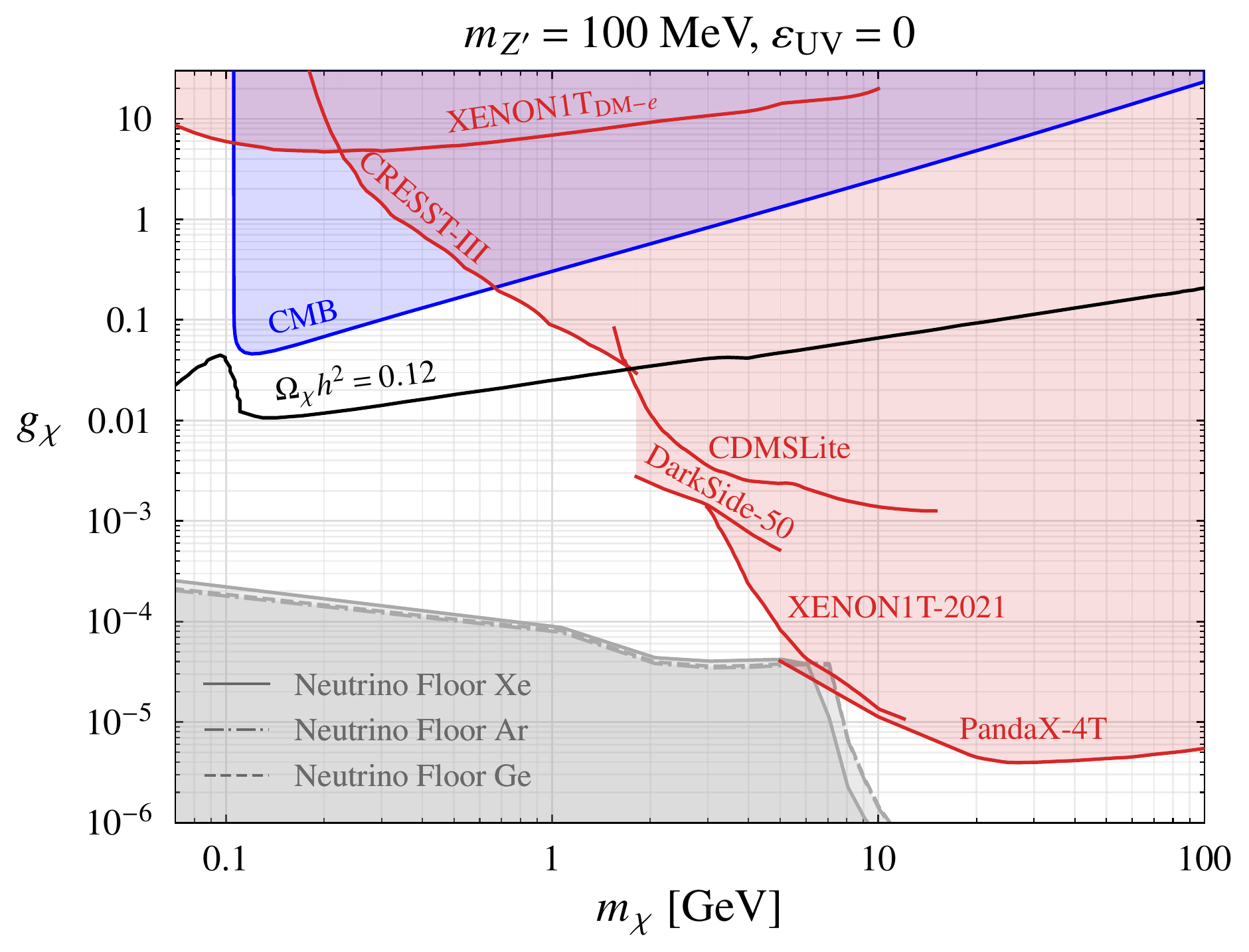}
\includegraphics[width=0.45\textwidth]{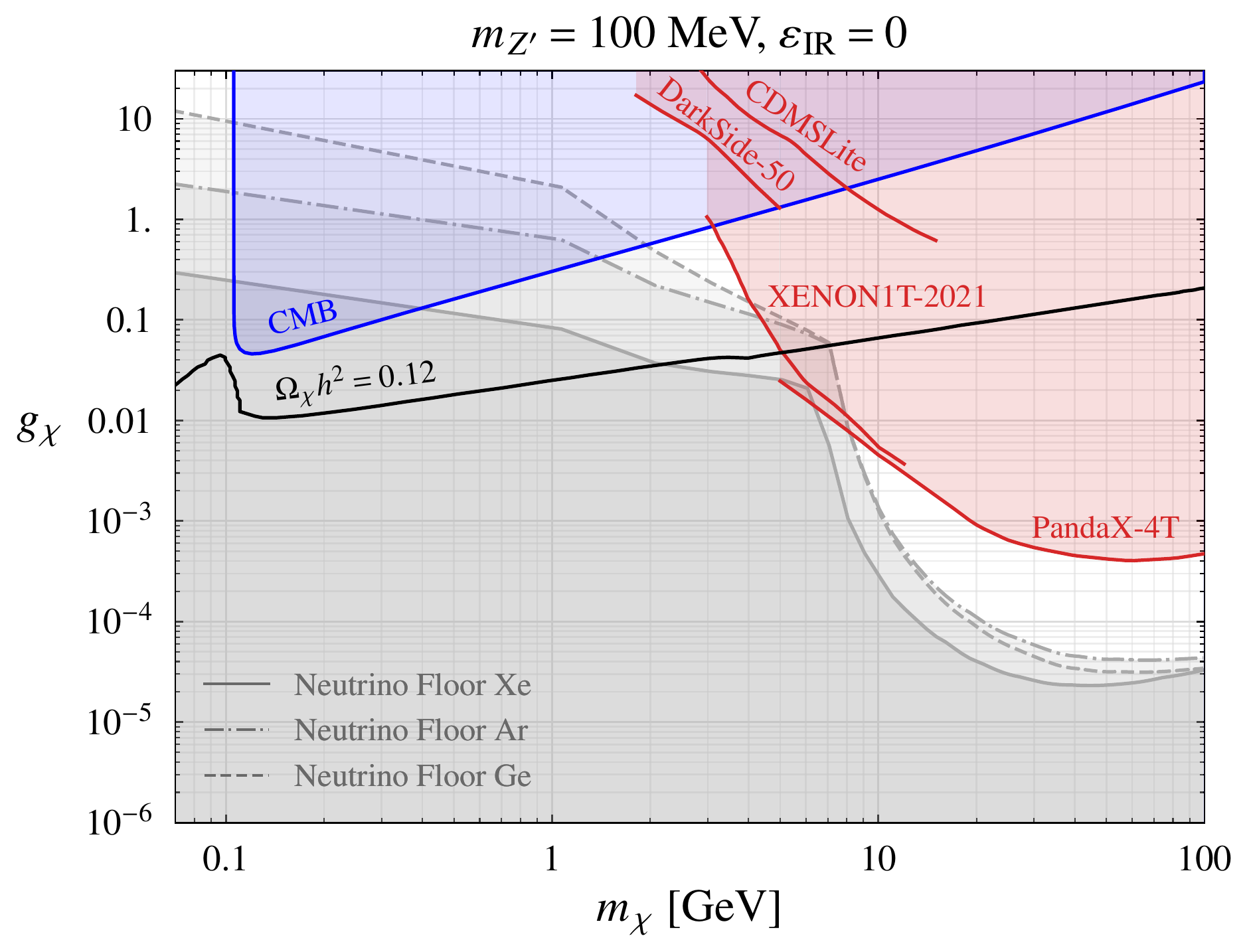}
\caption{
Thermal relic targets (black solid curves) and existing direct detection constraints (solid red curves) in the $L_\mu-L_\tau$ charged DM parameter space (coupling versus mass).
The direct detection limits depend on the choice of kinetic mixing and plots in the left and right columns correspond to $\varepsilon_{\rm UV}=0$ and $\varepsilon_{\rm IR}=0$, respectively.
We consider two benchmark $Z'$ masses, 20 MeV (plots in the first row) and 100 MeV (second row), with the corresponding gauge coupling $g_{\mu\tau}$ fixed to address the $(g-2)_\mu$ tension.
In addition, the blue shaded region is an indirect constraint on DM annihilation during the recombination epoch.
} \label{fig:rescaledlimits}
\end{figure}

To determine the upper bound on $g_\chi$ from direct detection experiments, we begin with the  $\varepsilon_{\rm UV}=0$ case. We first consider the heavy $Z'$ limit where Eq.~\eqref{eq:directdetectioncrosssection} applies. This allows us to directly translate the experimental upper limits on the DM-nucleon scattering cross section into limits on $g_{\chi}$ as a function of $m_\chi$ for given value of $m_{Z'}$.
We take into account existing results from CRESST-III~\cite{CRESST:2019jnq}, Darkside-50~\cite{Agnes:2018ves},  CDSMLite~\cite{SuperCDMS:2018gro}, XENON1T~\cite{XENON:2018voc,XENON:2019gfn,XENON:2020gfr}, and the very recent PandaX-4T result~\cite{Meng:2021mui}.
Next, to restore the momentum transfer dependence in the $Z^\prime$ propagator, we rescale this bound by a factor
\begin{equation}\label{eq:rescale0}
  \sqrt{\frac{R^{Q = 0}_{\varepsilon_\text{\sc uv=0}}}{R_{\varepsilon_\text{\sc uv=0}}}}
\end{equation}
where $R_{\varepsilon_\text{\sc uv=0}}$ is the direct detection event rate given in Eq.~\eqref{eq:rate}, evaluated with a kinetic mixing that satisfies a boundary condition $\varepsilon_{\rm UV} = \varepsilon_{\rm tot}(Q\to\infty)\to0$.
$R^{Q = 0}_{\varepsilon_\text{\sc uv=0}}$ corresponds to further replacing the factor $(Q^2+m_{Z'}^2)^2$ by $m_{Z'}^4$ in Eq.~\eqref{eq:rate}.
The quantity in Eq.~\eqref{eq:rescale0} is always larger than 1.

The resulting upper bounds on $g_\chi$ as a function of the DM mass $m_\chi$ are depicted in the top left and bottom left plots of Fig.~\ref{fig:rescaledlimits} for $m_{Z^\prime} = 10$ MeV and $m_{Z^\prime} = 100$ MeV, respectively. 
The corresponding coupling values $g_{\mu\tau} = 5.04\times10^{-4}$ and $9.5\times10^{-4}$ are used to fix the $(g-2)_\mu$ tension.
For the thermal DM target, we find that the DM mass is constrained by existing experiments to be less than $\sim 0.4$ (1) GeV for $m_{Z^\prime} = 10$ (100) MeV.
These results are roughly in agreement with the findings of Ref.~\cite{Asai:2020qlp}.
We also checked that the direct DM search limit using electron recoil~\cite{XENON:2016jmt} is too weak and does not place additional constraints. However, future experiments using electron recoils could potentially constrain the thermal relic target of the model for the $\varepsilon_\text{UV} = 0$ case \cite{Battaglieri:2017aum}.

However, it cannot be overemphasized that such a conclusion is derived by assuming a boundary condition $\varepsilon_\text{UV}=0$ for the kinetic mixing. As already stated earlier, the above upper bounds for $g_{\chi}$ and DM mass are not general. They could be modified significantly with other choices of the kinetic mixing. 
As a comparison, we consider the other boundary condition in Fig.~\ref{fig:running}, with $\varepsilon_{\rm IR}=0$ and explore the corresponding
DM phenomenology. We will comment on the more general choices near the end of this section.

\begin{figure}[t]
\centering
\includegraphics[width=0.45\textwidth]{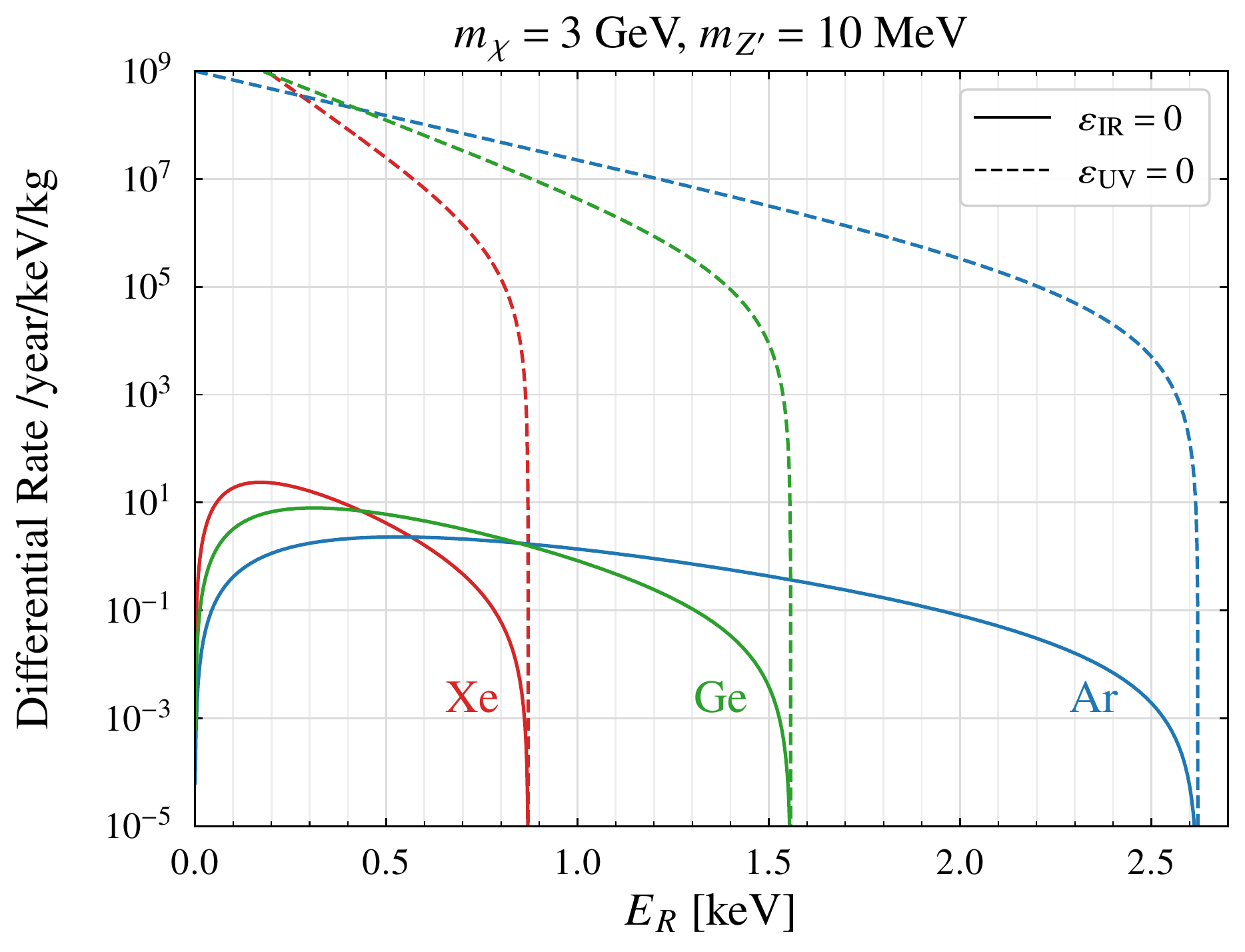}\hspace{0.5cm}
\includegraphics[width=0.445\textwidth]{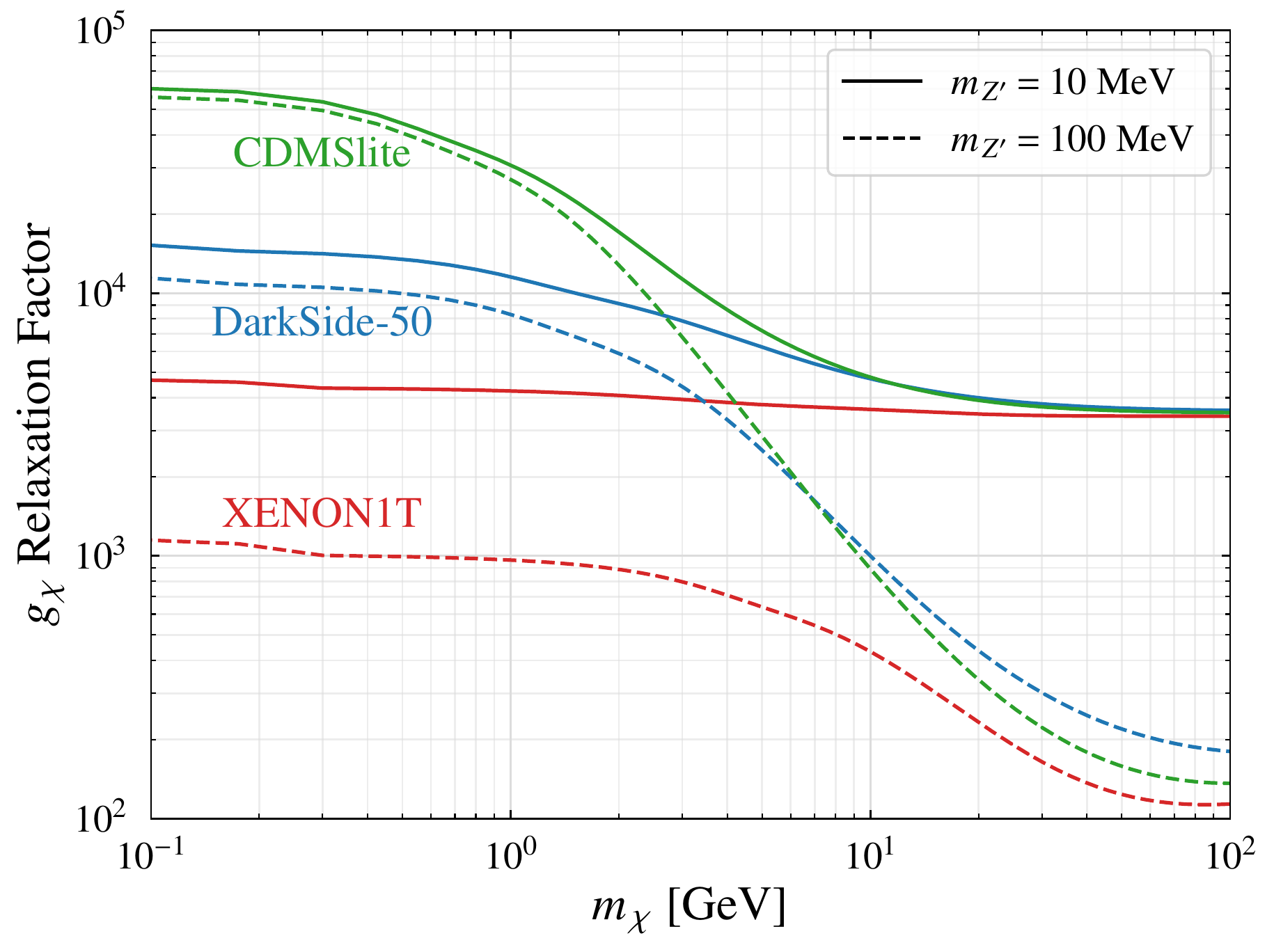}
\caption{{\it Left}: Differential scattering rates of DM with common detector materials under both considerations of the boundary conditions on $\varepsilon_{\rm tot}$, for $m_\chi = $ 3 GeV and $m_{Z'} = $ 10 MeV. It can be seen that for $\varepsilon_{\rm IR} = 0$ there is strong suppression in the associated rate $dR/dE_R$ relative to the $\varepsilon_{\rm UV} = 0$ case. 
{\it Right}: Relaxation factor for the dark coupling $g_\chi$ (defined in Eq.~\eqref{eq:momentumeffects}) as a function of $m_\chi$ for the experiments XENON1T (red curves), DarkSide-50 (blue curves), and CDMSLite (green curves). We use $m_{Z^\prime} = 10$ (solid curves) and $m_{Z^\prime}$ = 100 MeV (dashed curves) as benchmark values.  
The recoil energy thresholds of detectors we used are 2 keV for XENON1T, 0.6 keV for DarkSide, and 0.07 keV for CDMSLite.
}\label{fig:DifferentialRate}
\end{figure}

The momentum transfer dependence in the kinetic mixing affects the DM scattering rate through the $E_R$ integral of Eq.~\eqref{eq:rate}.
For direct detection, the typical momentum transfer $Q$ lies below the muon mass, thus we have approximately,  
\begin{equation}\label{eq:KE48}
\varepsilon_{\rm tot}(Q) \simeq 5\times10^{-4} g_{\mu\tau} \left(\frac{Q}{m_\mu}\right)^2 \ ,
\end{equation} 
in the $\varepsilon_{\rm IR}=0$ case, which is much smaller than $\varepsilon_{\rm tot}(Q) \simeq -g_{\mu\tau}/70$ for $\varepsilon_\text{UV}=0$.
Fig.~\ref{fig:DifferentialRate} (left) depicts the differential scattering rate $dR/dE_R$ in these two cases, for a set of model parameters, $m_{Z'}=10\,$MeV and $m_\chi=3\,$GeV.
The solid curves correspond to the case with $\varepsilon_{\rm IR}=0$ whereas the dashed curves correspond to $\varepsilon_{\rm UV}=0$.
Clearly, such a sizable suppression in the scattering rate implies much weaker direct detection limits for $\varepsilon_{\rm IR}=0$.

For more general choices of DM and $Z'$ masses, Fig.~\ref{fig:DifferentialRate} (right) shows the relaxation factor for the upper bound on $g_{\chi}$ in the $\varepsilon_{\rm IR}=0$ case compared to the $\varepsilon_{\rm UV}=0$ case, as a function of $m_\chi$.
The relaxation factor is defined as
\begin{equation}\label{eq:momentumeffects}
\sqrt{\frac{R_{\varepsilon_\text{\sc uv=0}}}{R_{\varepsilon_\text{\sc ir=0}}}}
\end{equation}
where $R$ is the direct detection event rate given in Eq.~\eqref{eq:rate} evaluated with a boundary condition for $\varepsilon_{\rm tot}(Q)$.
In this comparison, the upper bound on $g_{\chi}$ can be relaxed by as large as 4 orders of magnitude in the case of a GeV scale DM scattering off of an argon target.

We are now able to derive the upper bound on $g_\chi$ in the $\varepsilon_{\rm IR}=0$ scenario by rescaling the limits in the left panels of Fig.~\ref{fig:rescaledlimits} with the relaxation factor, Eq.~\eqref{eq:momentumeffects}.
The results are shown in the top right and bottom right plots of Fig.~\ref{fig:rescaledlimits}. 
We find the direct detection constraints indeed weaken substantially in this case. 
At the same time, the neutrino floors also rescale along with the direct detection limits.
The neutrino floor for different target materials (Ar, Xe, Ge) rescale by different amount because the rescaling factor Eq.~\eqref{eq:momentumeffects} is detector dependent.
Interestingly, this results in a new interplay with the thermal DM target.
For the $m_{Z'}=100\,$MeV case, the portion of the thermal relic curve not excluded by existing direct detection constraints is fully buried under the neutrino floor.
This is because existing limits from XENON1T and PandaX-4T already are strong enough to touch the neutrino floor around 5--6 GeV DM mass, 
whereas the thermal relic curve happens to pass across this region.
In contrast, for $m_{Z'}=10$\,MeV case, we find there is still a window of viable thermal DM mass (around $\sim1-4$\,GeV) that survives above the neutrino floors.
Such a new window (associated with a light $Z'$) serves as a nice target of upcoming direct detection experiments, such as DarkSide-LowMass~\cite{DarkSideLM, DarkSideLM-talk} and SuperCDMS-Ge~\cite{SuperCDMS:2016wui} that will be running at SNOLAB (see Fig.~\ref{fig:rescaledlimits_zoomin} below).

\begin{figure}[t]
\centering
\includegraphics[width=0.5\textwidth]{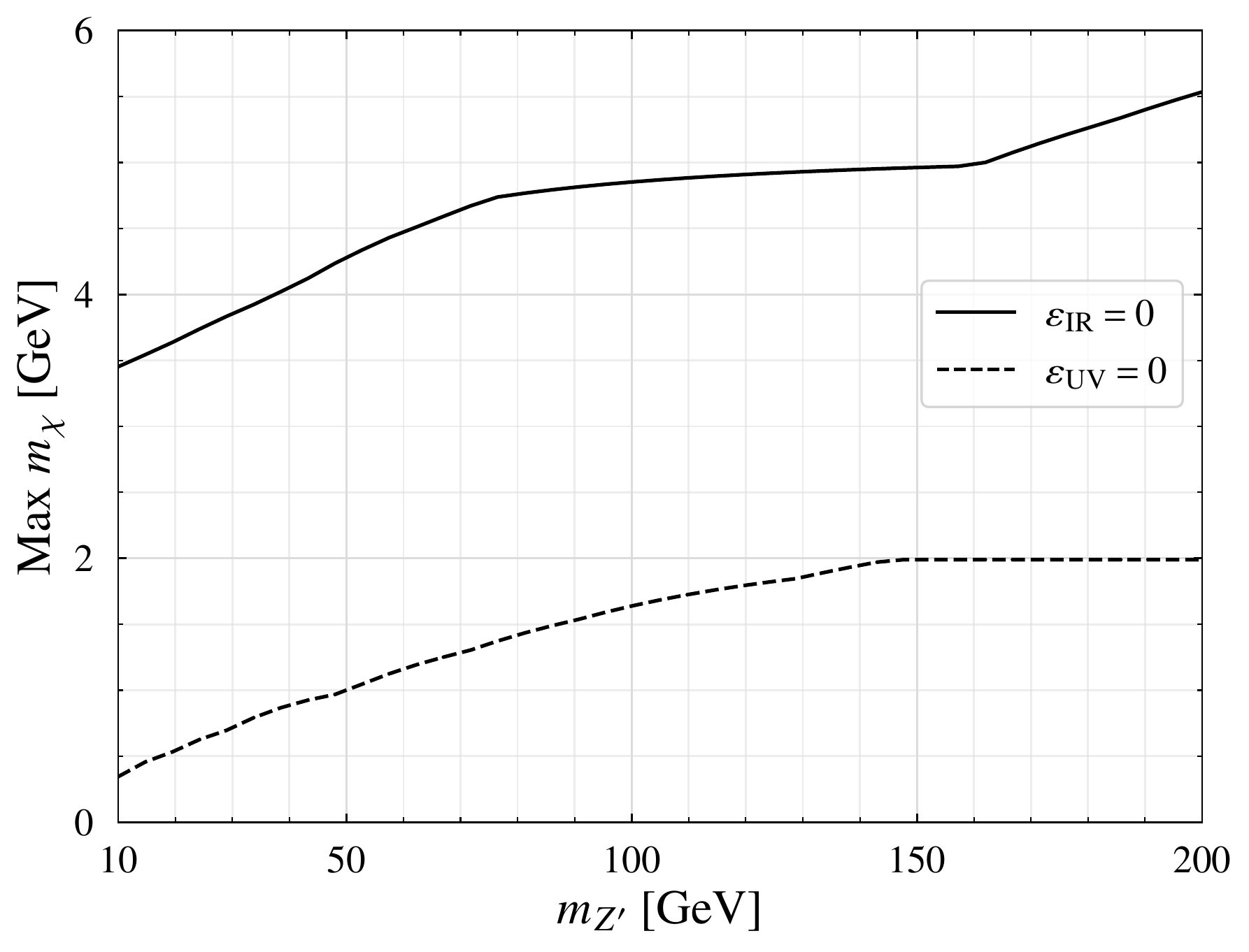}
\caption{Upper bounds on the DM mass in the gauged $U(1)_{L_\mu - L_\tau}$ model for two scenarios of the kinetic mixing,
$\varepsilon_{\rm UV}=0$ (dashed curve) and $\varepsilon_{\rm IR}=0$ (solid curve). }\label{fig:maxmass}
\end{figure}

The upper bound on the DM mass $m_\chi$ can be determined as the intersection point of the relic density curve with the edges of direct detection excluded region. Fig.~\ref{fig:maxmass} quantifies this upper bound by scanning the ${Z'}$ mass for both the $\varepsilon_{\rm UV} = 0$ and $\varepsilon_{\rm IR}=0$ scenarios. 
We find that the upper bound on $m_\chi$ can be relaxed up to $\sim 6$\,GeV with $\varepsilon_{\rm IR}=0$, in contrast to $m_\chi\lesssim 1$\,GeV in the $\varepsilon_{\rm UV} = 0$ case.
The mass of $Z'$ has an effect on these results, which could be inferred from Fig.~\ref{fig:rescaledlimits}.
In the $\varepsilon_{\rm UV} = 0$ case, direct detection limit on the thermal target curve is mainly set by CRESST-III, which gets stronger for lighter $Z'$ (the propagator effect).
As a result, the upper bound on $m_\chi$ reduces for lighter $Z'$, which agrees with the behavior of the dashed curve in Fig.~\ref{fig:maxmass}.
In the $\varepsilon_{\rm IR} = 0$ case, we observe similar behavior but it is PandaX-4T and XENON1T that dictate the upper bound on $m_\chi$.


\subsection{Novel recoil energy spectrum}

\begin{figure}[t]  
\includegraphics[width=0.7\textwidth]{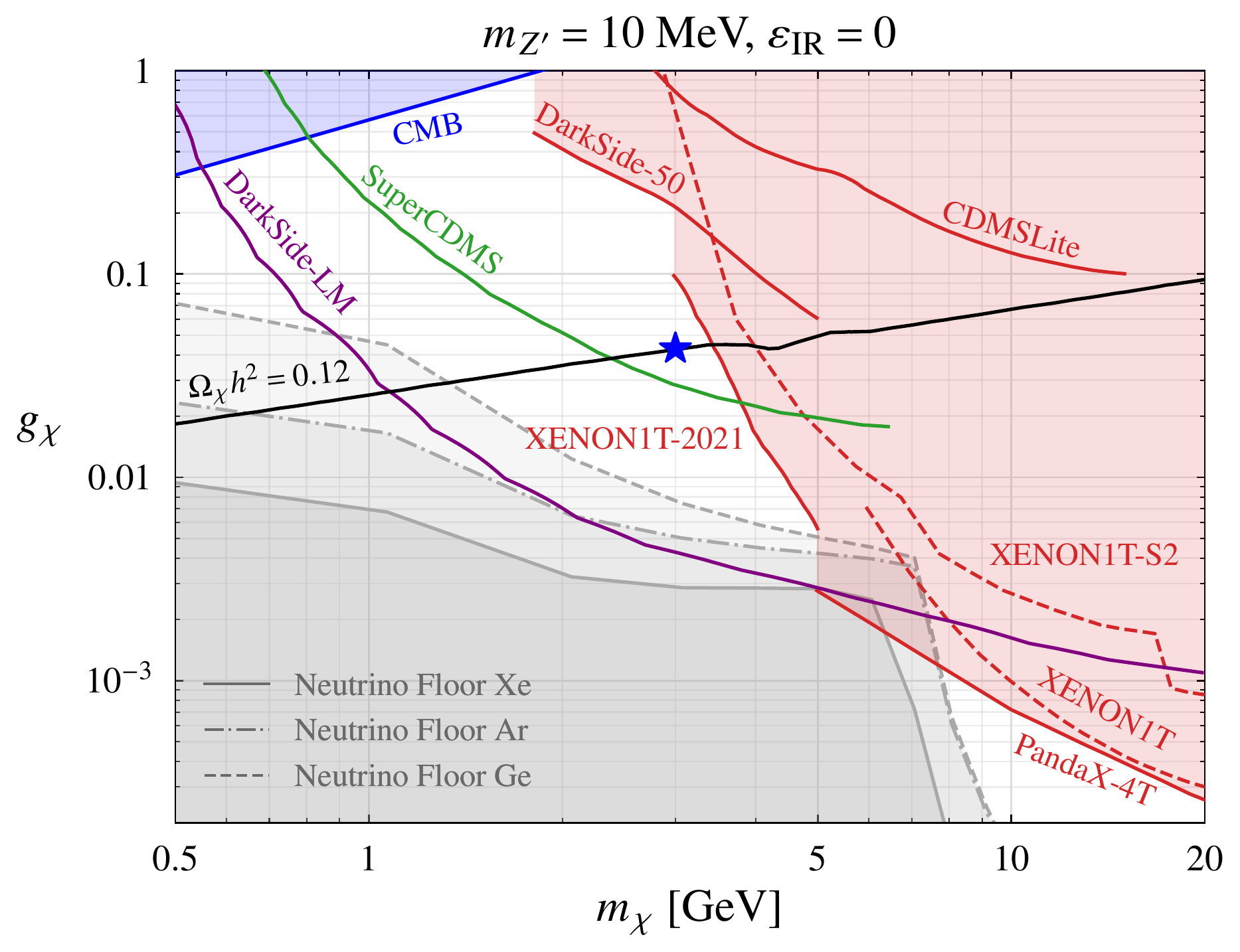}
\caption{
A zoomed-in version of Fig.~\ref{fig:rescaledlimits} (right) for the case of $m_{Z'}=10\,$MeV. Again, the black curve is the the thermal DM target in the light of $(g-2)_\mu$.
The red shaded regions are existing constraints. Future reaches by the DarkSide-LowMass and SuperCDMS-Ge experiments are shown as the purple and green curves, respectively.
} \label{fig:rescaledlimits_zoomin}
\end{figure}

As discussed above, for a kinetic mixing satisfying $\varepsilon_{\rm IR}=0$ together with a light $Z'$ boson, 
we open up a new window of thermal DM mass above GeV scale that is consistent with existing direct detection limits
and can be probed by the upcoming experiments.
To highlight the interplay, we show Fig.~\ref{fig:rescaledlimits_zoomin}, as a zoomed-in version of Fig.~\ref{fig:rescaledlimits} (right) for the case of $m_{Z'}=10\,$MeV.
On top of it, we include the future sensitivities of  DarkSide-LowMass and SuperCDMS-Ge experiments as purple and green curves, respectively.
Clearly, they stand in very good position to cover the remaining DM target above the neutrino floor. Note, that we show additional existing constraints from one tonne-year exposer of XENON1T \cite{XENON:2018voc} (dashed red curve labelled ``XENON1T'') and the XENON1T results from secondary scintillation (S2) only analysis \cite{XENON:2019gfn} (dashed red curve labelled ``XENON1T-S2'').

In this subsection, we derive another important consequence of the momentum dependence in the kinetic mixing,
by exploring the recoil energy spectra of DM scattering in the DarkSide-LowMass and SuperCDMS experiments. 
To proceed, we choose a set of benchmark parameters with 
\begin{equation}\label{eq:benchmarkpoint}
m_\chi=3\, {\rm GeV}, \quad m_{Z'} = 10\, {\rm MeV}, \quad g_\chi=0.04, \quad g_{\mu\tau} = 5.04\times10^{-4} \ , 
\end{equation}
which corresponds to the blue five-star in Fig.~\ref{fig:rescaledlimits_zoomin}.

Fig.~\ref{fig:DifferentialRate} already shows that the differential DM scattering rate $dR/dE_R$ is strongly suppressed by the small total kinetic mixing evaluated at low $Q$, in the case of $\varepsilon_{\rm IR}=0$. Eq.~\eqref{eq:KE48} further tells us that such a suppression effect is the strongest at $Q=0$ where $\varepsilon_{\rm tot}(Q)$ vanishes completely.
For higher recoil energies, the differential rate first increases following the behavior of $\varepsilon_{\rm tot}$ and eventually shuts off when the scattering runs out of phase space.
This interplay predicts a peak in the $dR/dE_R$ curve at intermediate recoil energies. 

\begin{figure}[t]  
\includegraphics[width=0.425\textwidth]{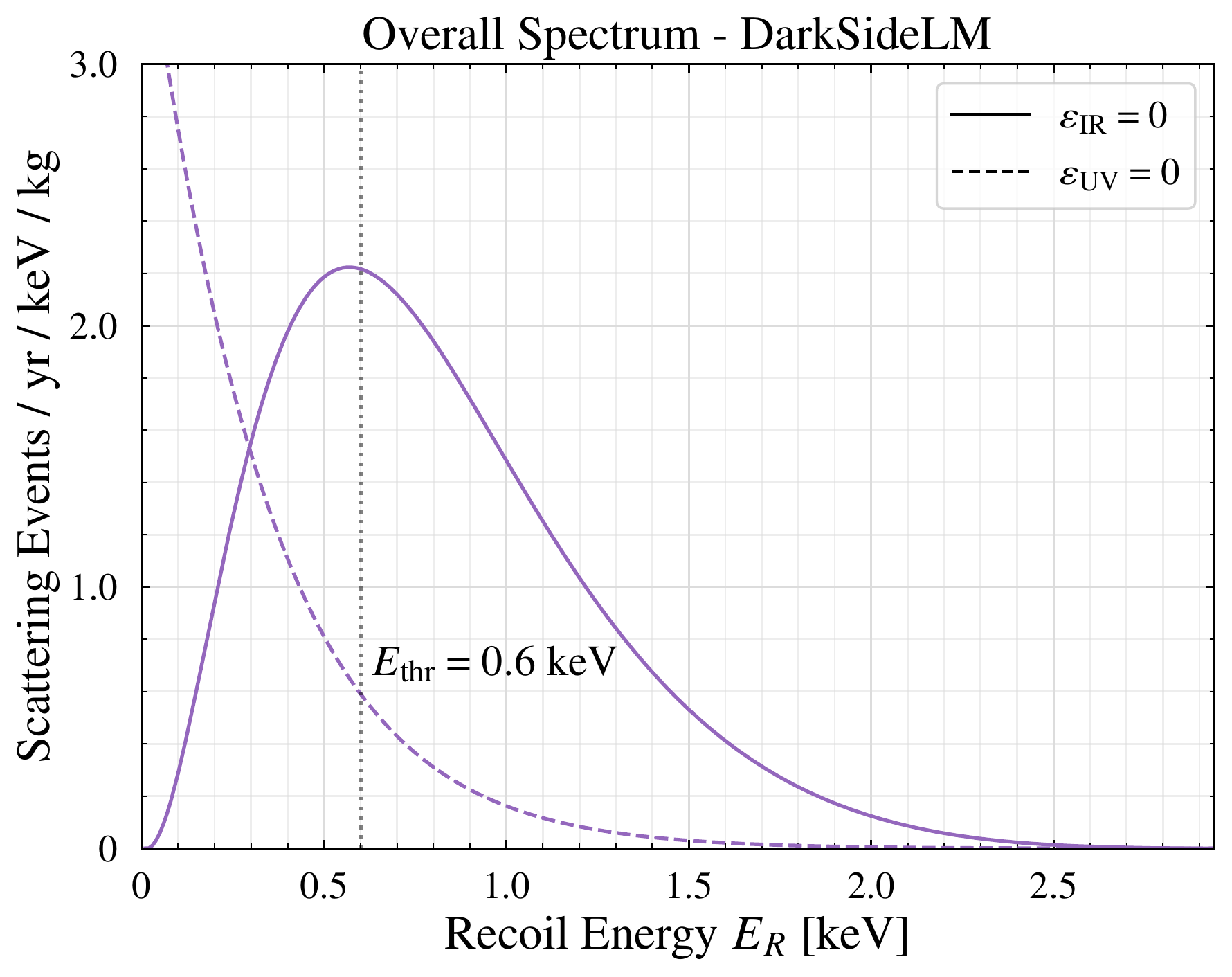}\hspace{1cm}
\includegraphics[width=0.425\textwidth]{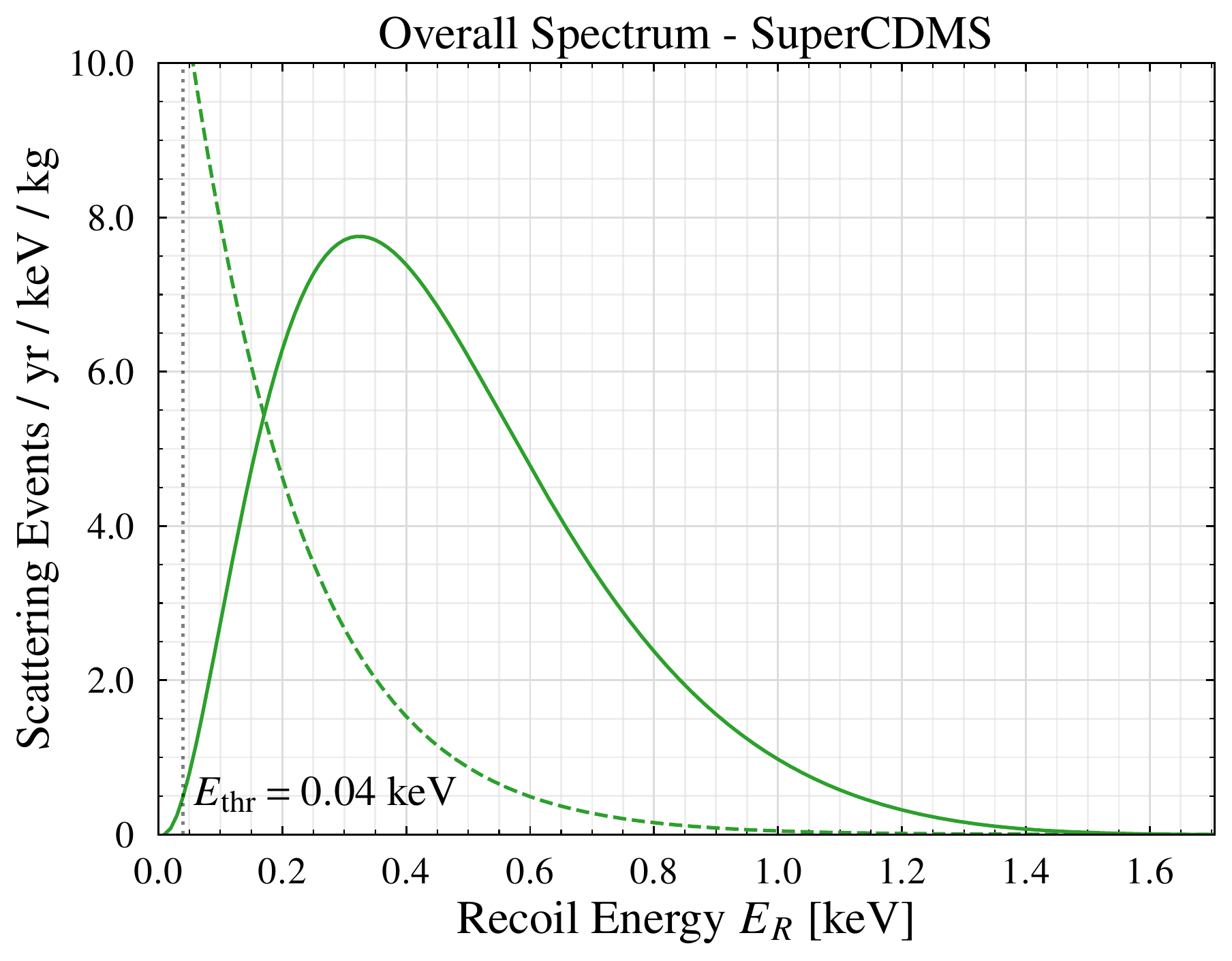}\\[4ex]
\includegraphics[width=0.425\textwidth]{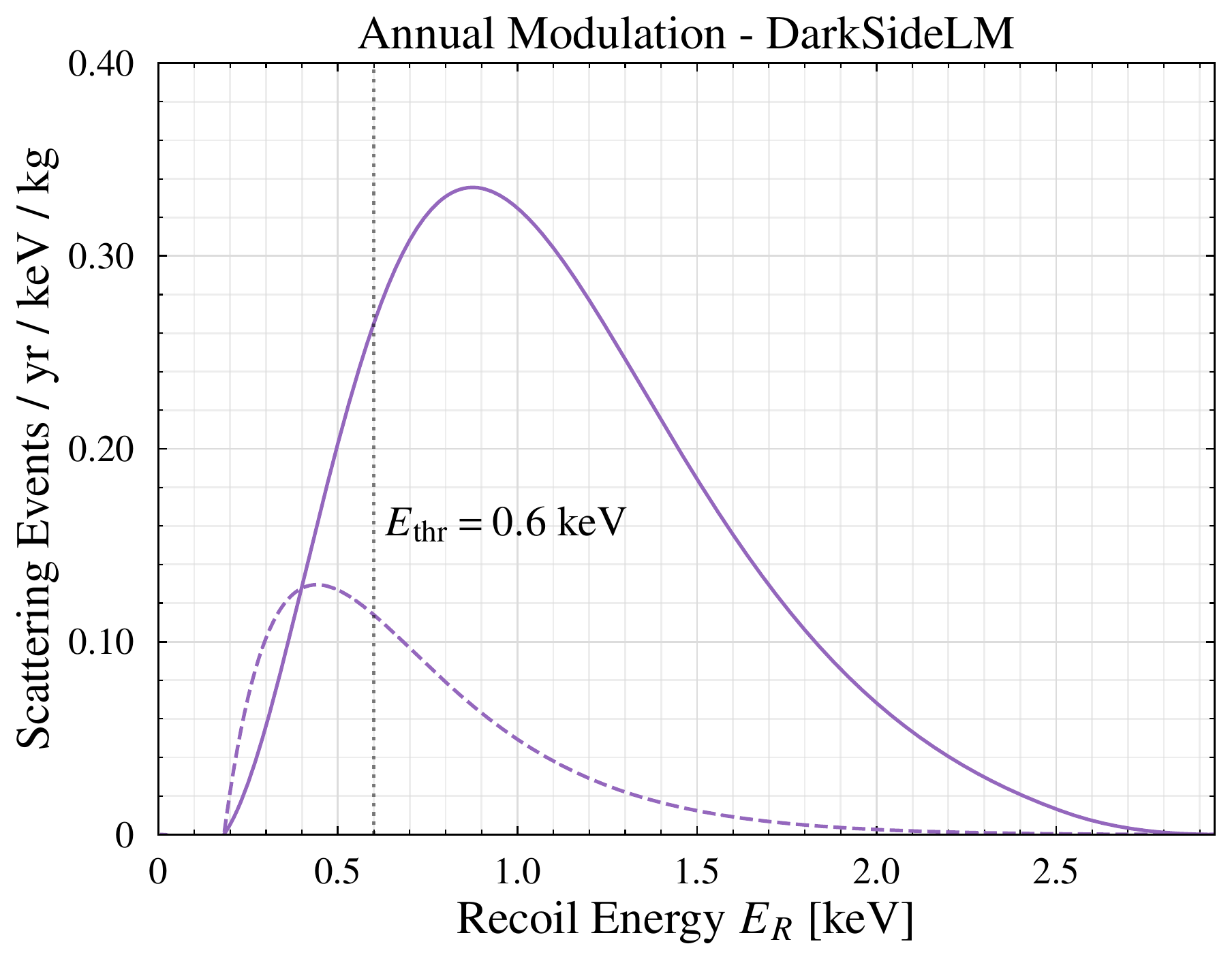}\hspace{1cm}
\includegraphics[width=0.425\textwidth]{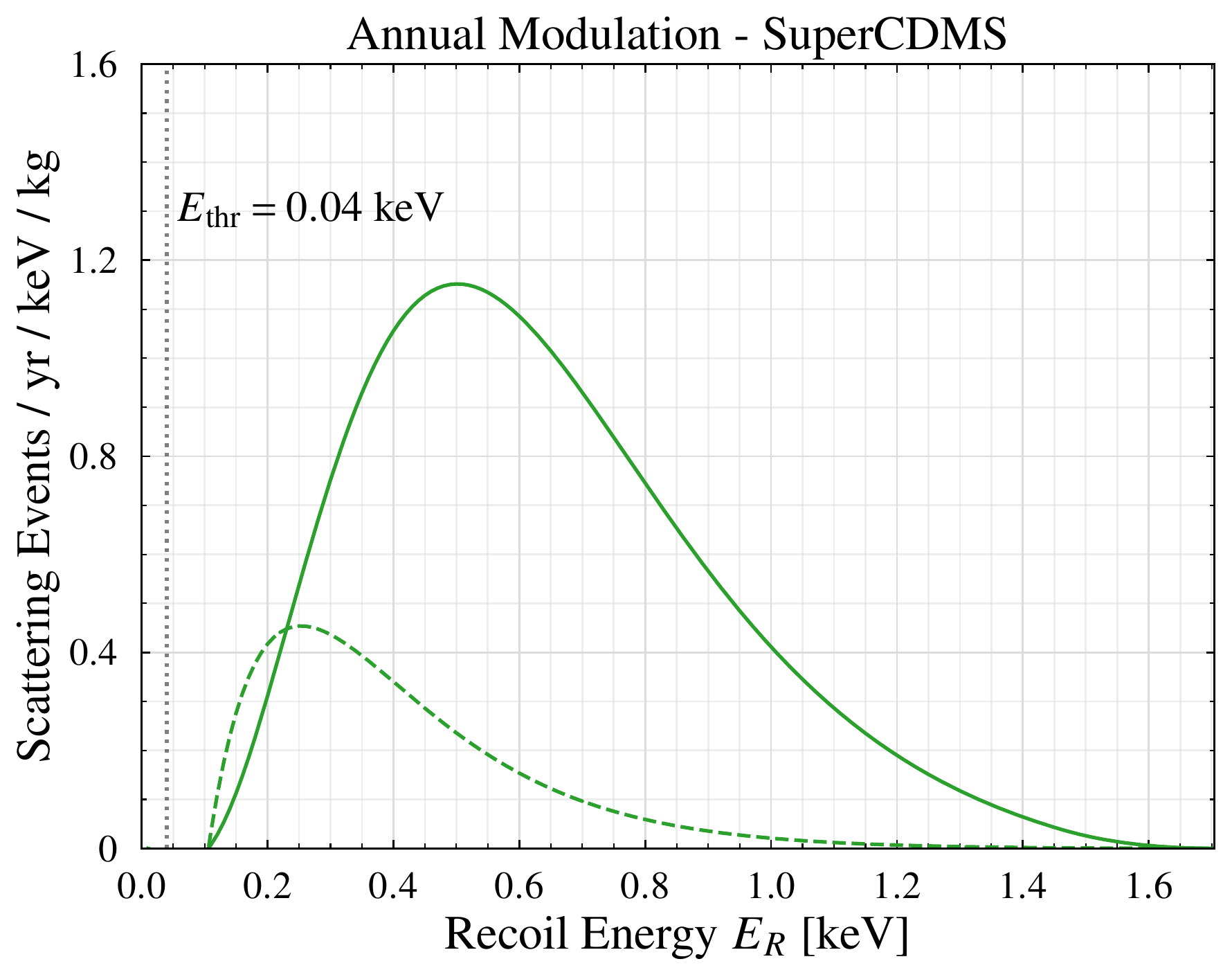}
\caption{Averaged and annual modulated DM recoil spectra at the future DarkSide-LM and SuperCDMS experiments for two scenarios of the kinetic mixing,
$\varepsilon_{\rm UV}=0$ (dashed curves) and $\varepsilon_{\rm IR}=0$  (solid curves).
The normalization of the $\varepsilon_{\rm UV}=0$ curves is chosen so that it gives the same total scattering rate as the $\varepsilon_{\rm IR}=0$ case.
} \label{fig:modulation}
\end{figure}

All the above features are captured by the solid curves in the first row of Fig.~\ref{fig:modulation}.
The shape of recoil spectrum differs from normal elastic scattering with constant couplings (i.e. the $\varepsilon_{\rm UV}=0$ case) whose differential rate $dR/dE_R$ peaks at zero recoil energy,
as shown by the dashed curves in Fig.~\ref{fig:modulation}. Instead, for $\varepsilon_{\rm IR}=0$ the recoil spectrum peaks at a nonzero recoil energy. 
Such a novel shape of the recoil energy spectrum can serve as an important ``smoking-gun'' signal for upcoming direct detection experimental searches, especially those featuring low detector thresholds, DarkSide-LowMass ($E_{\rm th}=0.6$\,keV)~\cite{DarkSideLM} and SuperCDMS-Ge ($E_{\rm th}=40$\,eV)~\cite{SuperCDMS:2016wui}.

It is worth emphasizing again that for the benchmark parameters considered in Eq.~\eqref{eq:benchmarkpoint}, 
only the $\varepsilon_{\rm IR}=0$ case allows for a simultaneous explanation of $(g-2)_\mu$ and thermal DM relic abundance without being excluded by existing direct detection limits.
The $\varepsilon_{\rm UV}=0$ case fails to do so. The dashed curves in Fig.~\ref{fig:modulation} are depicted with arbitrary normalization for comparison and could only work if one of the assumptions is relaxed.

We also derive the annual modulated recoil spectrum in the two types of detectors, which is given by the difference when the parameter $v_\odot$ in DM velocity distribution (see Eq.~\eqref{eq:rate}) is replaced by $v_\odot \pm v_\oplus$, where $v_\oplus \simeq 30\,{\rm km/s}$ is the velocity of the earth around the sun.
The annual modulation spectra are displayed in the bottom row of Fig.~\ref{fig:modulation}. 
For $\varepsilon_{\rm IR}=0$, they inherit a similar peaked shape like the averaged spectrum, driven by the kinetic mixing effect.
The recoil energies associated with the peaks of the spectra are much higher for $\varepsilon_{\rm IR}=0$ than the case with $\varepsilon_{\rm UV}=0$ where DM scatters via nearly constant couplings.
This feature adds another handle to the smoking-gun signature. 

To summarize, the DM recoil energy spectrum encodes important information of $L_\mu-L_\tau$ charged DM and the nature of kinetic mixing.
It serves as a well motivated target for the upcoming search at DarkSide-LM and SuperCDMS experiments.
For general kinetic mixing scenarios beyond the $\varepsilon_{\rm IR}=0$, $\varepsilon_{\rm UV}=0$ cases, the novel $E_R$ (or $Q$) dependence in the kinetic mixing could still be 
tested with a precision measurement of the DM recoil energy spectrum at low threshold detectors.
Having multiple detectors and target materials will be helpful for testing the $E_R$ dependence.

\subsection{Other Constraints}

\subsubsection{Indirect detection using CMB and neutrinos} 

Going beyond direct detection, we comment on other approaches to probe the $L_\mu - L_\tau$ charged DM. 
For $m_\chi > m_{\mu,\tau}$ DM can annihilate into these visible particles which leads to indirect signals.
The annihilation cross section is given by the third line of Eq.~\eqref{annihilationchannels}. 
Energy injection due to DM annihilation during the recombination era is constrained by the observed cosmic microwave background (CMB) spectrum~\cite{Slatyer:2015jla, Kawasaki:2021etm}.
Assuming $\chi$, $\bar\chi$ make up 100\% of DM in the universe, 
the current CMB exclusion of the model parameter space is shown by the blue shaded regions in Fig.~\ref{fig:rescaledlimits}.
We find it is not strong enough to touch the thermal relic curve, primarily because $\chi\bar\chi \to \mu^+\mu^-$ and $\tau^+\tau^-$ are subdominant annihilation channels.

There are also indirect searches for galactic DM annihilating into neutrinos, which occurs through $\chi\bar\chi \to Z'Z'$ followed by $Z'\to \nu\bar\nu$ in the parameter space we consider.
The corresponding constraints are yet to reach the thermal target~\cite{Arguelles:2019ouk}.
These indirect constraints are robust and do not depend on assumptions of the kinetic mixing.

\subsubsection{Dark matter self interaction} 

The DM model setup considered here features a light mediator $Z'$, which allows DM to self interact.
For GeV scale thermal DM and the mass range of $Z'$ of interest to $(g-2)_\mu$, it is sufficient to estimate the self interaction using the Born approximation~\cite{Feng:2009hw}.
The momentum transfer cross section satisfies 
\begin{equation}
\sigma_T/m_\chi \simeq \frac{g_\chi^4 m_\chi}{4\pi m_{Z'}^4} \simeq 0.02\,{\rm cm^2/g} \left( \frac{m_\chi}{\rm 3\,GeV} \right) \left( \frac{10\,\rm MeV}{m_{Z'}} \right)^4 \ .
\end{equation}
This implies that the model parameter space explored in this work is compatible with DM self-interaction constraints such as the bullet cluster~\cite{Randall:2008ppe}.

\section{Conclusion}\label{sec:conclusion}

In this work, we explore neutrino and DM phenomenology in the gauged $U(1)_{L_\mu-L_\tau}$ model that addresses the $(g-2)_\mu$ tension.
A number of experimental probes are controlled by the kinetic mixing between the new gauge boson $Z'$ and the photon, which depends on the bare kinetic mixing $\varepsilon_0$ as a free parameter.
In general, the effective kinetic mixing in a process depends on the typical momentum transfer.
By comparing different scenarios for the kinetic mixing, we find that they predict drastically different signal rates at low-energy neutrino experiments.
More concretely we find that, the constraint on neutrino-electron scattering set by the Borexino experiment, constraints on neutrino-nucleus scattering set by the COHERENT experiments, as well as the white dwarf cooling constraint, all can be substantially weakened if the asymptotic value of the kinetic mixing vanishes at zero momentum transfer. 
In this case, they do not impose any constraint on the $(g-2)_\mu$ favored target parameter space of the model.

As a further step, we explore the impact of the kinetic mixing effect on $L_\mu-L_\tau$ charged DM.
We make the minimal assumption where the DM is a Dirac fermion and obtain the correct relic abundance by annihilating into $Z'$ bosons.
We find direct detection constraints can also be relaxed with the same choice of kinetic mixing that allows the above neutrino constraints to be lifted.
This helps to open up a mass window for the DM above the GeV scale which can be further explored by future low-threshold DM direct searches including DarkSide and SuperCDMS.
The interplay with momentum dependence in the kinetic mixing leads to novel recoil energy spectrum in DM scattering. 

Our work demonstrates the phenomenological importance of the kinetic mixing in gauged $U(1)_{L_\mu-L_\tau}$ extension of the Standard Model motivated by the muon $g-2$ anomaly.
A joint effort of future neutrino and DM experiments and precision spectral measurement using low threshold detectors will be the key to test such a theory.

\bigskip
\section*{Acknowledgment}
We thank Ning Zhou for useful discussions. 
This work is supported by the Arthur B. McDonald Canadian Astroparticle Physics Research Institute.

\appendix

\section{Statistical Analysis for CE$\nu$NS}\label{sec:stats}

In this appendix, we describe the statistical analysis used to derive the constraints from CE$\nu$NS shown in Fig.~\ref{fig:neutrino}. We follow closely the discussion in \cite{Banerjee:2021laz}. To determine upper limits on the parameter space of the $U(1)_{L_\mu - L_\tau}$ model we use a chi-squared test given by
\begin{equation}\label{eq:chisquare}
\chi^2 = \underset{\alpha}{\text{Min}} \Bigg[ \frac{\big(N_\text{exp} - (1 + \alpha) N_\text{th}\big)^2}{\sigma^2_\text{stat}} + \bigg(\frac{\alpha}{\sigma_\alpha}\bigg)^2 \Bigg]
\end{equation}
where $N_\text{exp}$ is the number of CE$\nu$NS events observed by a given experiment, $N_\text{th}$ is the predicted number of CE$\nu$NS events in the model given by Eq.~\eqref{eq:Ncevns}, $\sigma_\text{stat}$ is the statistical uncertainty in the $N_\text{exp}$, and $\alpha$ is an overall normalization factor that accounts for the systematic uncertainty $\sigma_\alpha$. 

For the COHERENT CsI and CENNS-10 experiments, the statistical uncertainty in the number of observed CE$\nu$NS events  is given by $\sigma_\text{stat} = \sqrt{N_\text{exp} + N_\text{SS} + N_\text{BRN} + N_\text{NIN}}$, where $N_\text{SS}$ is the observed number of steady-state background events, $N_\text{BRN}$ is the number of beam-related neutron events, and $N_\text{NIN}$ is the number of neutrino-induced neutrons. The systematic uncertainties for CsI and argon experiments are $\sigma_\alpha = 0.13$ and $\sigma_\alpha = 0.085$, respectively. Note, that in our chi-square test we use the fit results of Analysis A of the CENNS-10 data \cite{COHERENT:2020iec}.

For future liquid argon experiments, we assume that the observed number of events is consistent with SM predictions, and use a similar chi-square test as in Eq.~\eqref{eq:chisquare} with $N_\text{exp}$ replaced by the number of CE$\nu$NS events predicted in the SM $N_\text{SM}$. The statistical uncertainty is given by $\sigma_\text{stat} = \sqrt{N_\text{SM} + N_\text{bkg}}$ where we assume that $N_\text{bkg} = N_\text{SM}/ 10$. In addition, we assume a systematic uncertainty $\sigma_\alpha = 0.05$.

After minimizing Eq.~\eqref{eq:chisquare} with respect to $\alpha$, the 95\% exclusion limits from CE$\nu$NS is found by requiring that $\Delta \chi^2 = \chi^2 - \chi^2_\text{SM} = 5.99$ ($\chi^2_\text{SM}$ is found by making the replacement $N_\text{th} \to N_\text{SM}$ in Eq.~\eqref{eq:chisquare}). This requirement will give us a value of $N_\text{th}$ that can then be used to find the upper limit on the coupling $g_{\mu\tau}$, shown in the bottom row Fig.~\ref{fig:neutrino}.

\begin{figure*}[t]  
\includegraphics[width=0.45\textwidth]{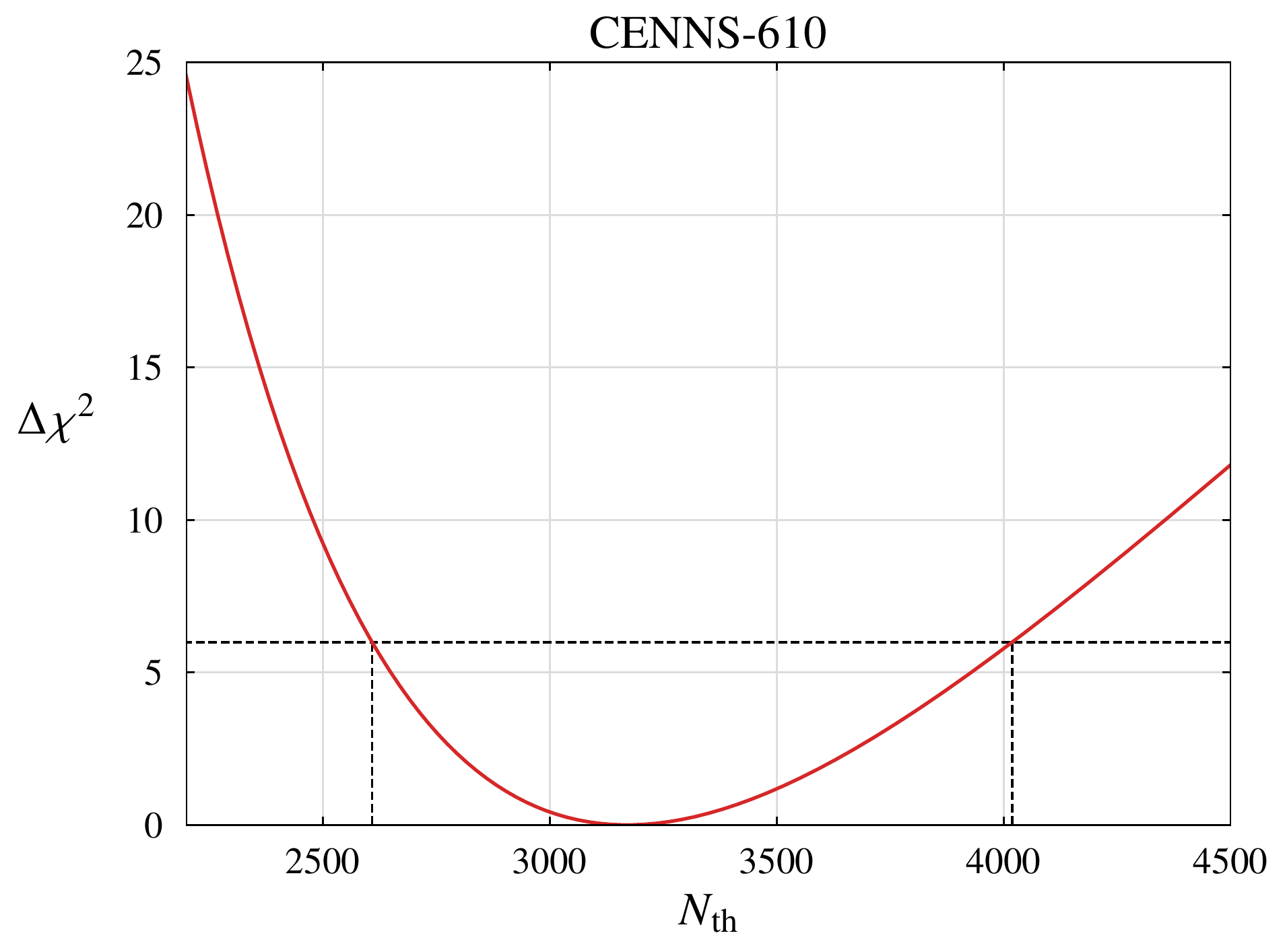}~~
\includegraphics[width=0.45\textwidth]{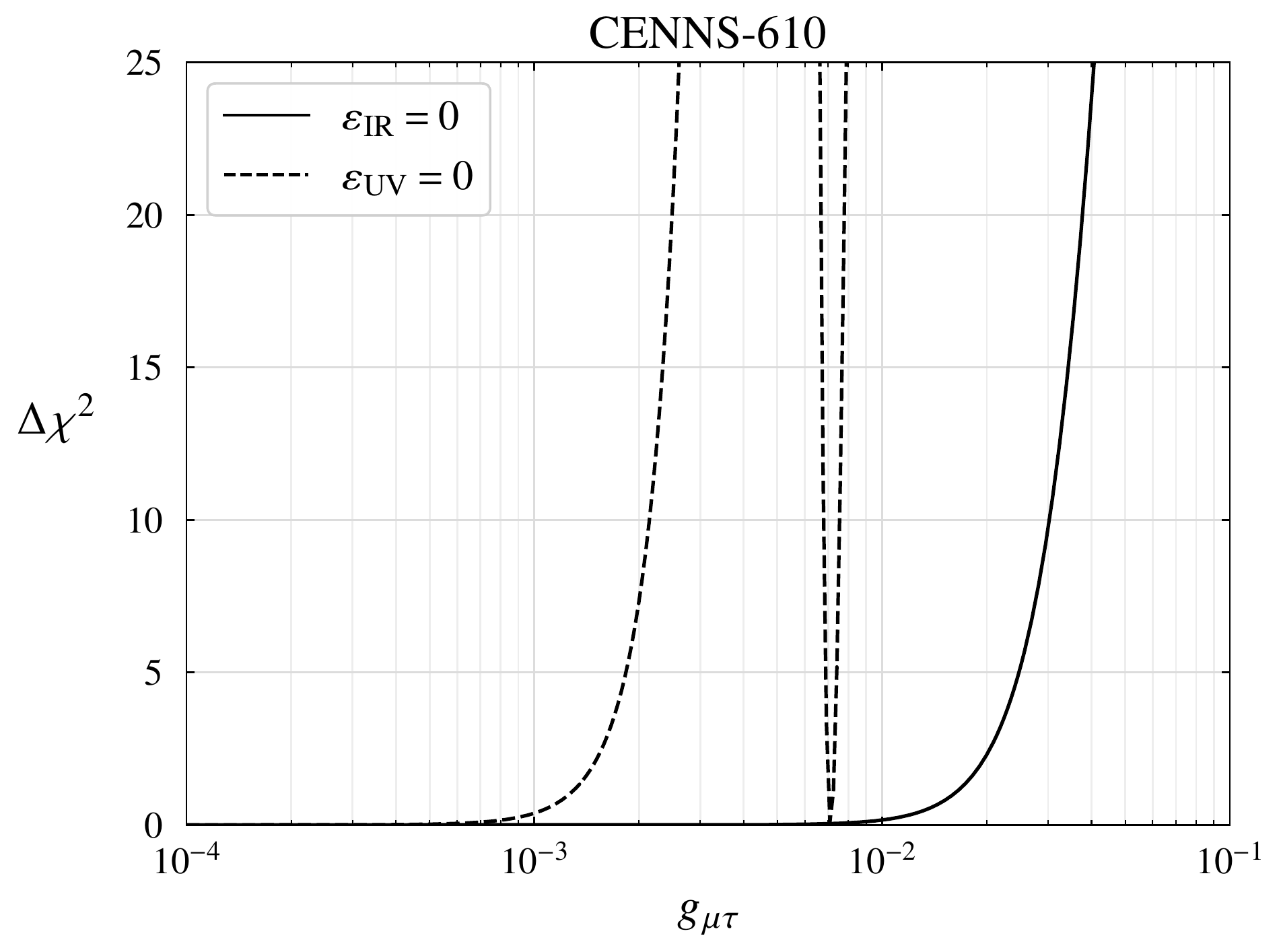}\\

\caption{$\Delta \chi^2$ as a function of theory prediction (left) and gauge coupling $g_{\mu\tau}$ (right) for a CE$\nu$NS experiment in the gauged $U(1)_{L_\mu-L_\tau}$ model. We use the proposed CENNS-610 experiment for illustration and $m_{Z^\prime} = 100$ MeV.} \label{fig:chisquare}
\end{figure*}

Note that this limit depends on the sign of $g_{\mu\tau}$ for the scenario with the boundary condition $\varepsilon_\text{UV} = 0$. For $g_{\mu\tau} > 0$ there is destructive interference between the SM and $Z^\prime$ contributions to CE$\nu$NS since from Eq.~\eqref{eq:Ncevns} we have 
\begin{equation}
\frac{d\sigma_{\nu_\mu N}}{dE_R} = \frac{d\sigma_{\bar\nu_\mu N}}{dE_R}  \propto \left| \frac{G_F Q_{\nu A}}{2} + \frac{g_{\mu\tau} \varepsilon_{\rm tot} Z e}{2 m_A E_R + m_{Z'}^2} \right|^2.
\end{equation}
In some regions of parameter space this destructive interference predicts a value of $N_{\text{CE}\nu\text{NS}}$ that is much lower than the experimental value. There will be a lower and upper limit on the value of $N_\text{th}$ that satisfies $\Delta \chi^2 = 5.99$, as shown in the left plot of Fig~\ref{fig:chisquare}. 

In practice, we find this only happens for the $\varepsilon_\text{UV}=0$ case. In the right plot of Fig.~\ref{fig:chisquare} we show $\Delta \chi^2$ as a function of $g_{\mu\tau}$ for $m_{Z^\prime} = 100$ MeV. For the boundary condition $\varepsilon_\text{IR}=0$ (solid curve) we find that there is a single value of $g_{\mu\tau}$ that satisfies $\Delta \chi^2 = 5.99$, corresponding to the maximum value of $N_\text{th}$ that is compatible with the expected number of observed events i.e this scenario only adds to the SM prediction. On the other hand, for $\varepsilon_\text{UV}=0$ (dashed curve) there are multiple values of $g_{\mu\tau}$ that satisfy $\Delta \chi^2 = 5.99$. The strongest constraint on $g_{\mu\tau}$ in this case is due to the destructive interference between the SM and $Z^\prime$, corresponding to the minimum value of $N_\text{th}$ that is compatible with the expected number of observed events.

{
\renewcommand{\arraystretch}{1.25}
\begin{table*}[t] 
\begin{ruledtabular}
\begin{tabular}{ c c c c c c  }
    Experiment & Mass [kg] & $E_\text{th}$ [keV$_{nr}$] & $N_\text{POT}$ & $r$ & $L ~[m]$ \\   \hline
        COHERENT CsI \cite{Akimov:2017ade, Akimov:2018vzs}& 14.6 & 4.25 & $1.76\times 10^{23}$& 0.08 & 19.3  \\  
        COHERENT LAr \cite{COHERENT:2020iec,COHERENT:2020ybo}& 24.4 & 20 & $1.38 \times 10^{23}$ & 0.09 & 27.5 \\   \hline
        CENNS-610 \cite{COHERENT:2019kwz} & 610 & 20 & $1.5 \times 10^{23}$ &  0.08& 28.4\\  
	ESS-10 \cite{Baxter:2019mcx}& 10 & 0.1 & $2.8 \times 10^{23}$  &  0.3 &  20\\   
	CCM \cite{CCM,CCM:2021leg} & 7000 & 10 & $0.177 \times 10^{23} $& 0.0425& 20\\ 

  \end{tabular}
\caption{Experimental parameters for current and future CE$\nu$NS experiments. For current CsI and LAr COHERENT experiments (first two rows) we quote the total number of protons-on-target during their full run time.; for future experiments we quote the number of protons-on-target $N_\text{POT}$  assuming one year of running.}\label{tab:coherent}
  \end{ruledtabular}
\end{table*}}

\bibliography{references}
\end{document}